\documentclass{mn2e} 
\usepackage{times}
\input{psfig.sty}

\newif\ifAMStwofonts

\title[A fundamental plane of black hole activity]
      {A fundamental plane of black hole activity} 

\author[Merloni, Heinz \& Di Matteo] {Andrea Merloni, Sebastian Heinz \&
Tiziana Di Matteo \\Max-Planck-Institut f\"ur Astrophysik,
Karl-Schwarzschild-Strasse 1, D-85741, Garching, Germany }

\date{}

\begin{document}

\maketitle

\label{firstpage}

\begin{abstract}
 
  We examine the disc--jet connection in stellar mass and supermassive
  black holes by investigating the properties of their compact emission in
  the X-ray and radio bands.  We compile a sample of $\sim$100 active
  galactic nuclei with measured masses, 5 GHz core emission, and 2-10 keV
  luminosities, together with 8 galactic black holes with a total of $\sim$
  50 simultaneous observations in the radio and X-ray bands. Using this
  sample, we study the correlations between the radio ($L_{\rm R}$) and
  the X-ray ($L_{\rm X}$) luminosity and the black hole mass ($M$).  We
  find that the radio luminosity is correlated with {\em both} $M$ and
  $L_{\rm X}$, at a highly significant level.  In particular, we show that
  the sources define a ``fundamental plane'' in the three-dimensional
  ($\log L_{\rm R},\log L_{\rm X},\log M$) space, given by $\log L_{\rm
  R}=(0.60^{+0.11}_{-0.11}) \log L_{\rm X} +(0.78^{+0.11}_{-0.09}) \log M +
  7.33^{+4.05}_{-4.07}$, with a substantial scatter of $\sigma_{\rm
  R}=0.88$. We compare our results to the theoretical relations between radio
  flux, black hole mass, and accretion rate derived by  Heinz \& Sunyaev
  (2003). Such relations depend only on the assumed accretion
  model and on the observed radio spectral index. Therefore,
  we are able to show that
  the X-ray emission from black holes accreting at less than a few per cent
  of the Eddington rate is unlikely to be produced by radiatively efficient
  accretion, and is marginally consistent with optically thin
  synchrotron emission from the jet. On the other hand, models for
  radiatively inefficient accretion flows seem to agree well with the data.

\end{abstract}


\section{Introduction}
The ultimate observational evidence of a celestial body being a black
hole comes from dynamical studies, by measuring the gravitational
influence of the central object on neighboring stars and gas, both in
binary systems and in the nuclei of galaxies. However, there are a
number of distinctive signatures of black hole-powered activity that
are usually regarded as proxy of black hole existence.  Relativistic
jets emitting synchrotron radiation in the radio band are one such
signature, the second most common being the presence of strong, compact
power-law X-ray emission commonly associated with the inner part of an
accretion flow\footnote{For the case of stellar mass objects, these
  signatures are common to most neutron stars, too. In this cases,
  only dynamical studies can provide a proof of the black hole nature
  of the source.}.  Indeed, as a general property, accretion onto compact
objects and the launch of relativistic outflows/jets seem to be
correlated (or {\em symbiotic}, Falcke \& Biermann, 1995) phenomena
\cite{bbr84,rs91}. This implies that, at some level, a correlation
between jet and disc flux is unavoidable.

Observationally, jet morphologies and spectral properties of both radio and
X-ray cores are remarkably similar in the case of black holes of stellar
mass (galactic black holes, hereafter GBH) and of their supermassive
counterparts  in the nuclei of galaxies (hereafter SMBH).  If jets are
launched in the innermost parts of the accretion flows, as commonly
assumed, then these similarities suggest that it should be possible to
understand the physics of both black hole accretion and jet production by
studying all those systems as a {\it single} class.
To this end, radiation emerging at radio and (hard) X-ray frequencies 
is the most direct
probe of the immediate vicinity of the black hole: effects of extinction
are unimportant in the radio band, while in the X-rays, where a large
fraction of the accretion energy is radiated, 
such effects can be accounted for with
good enough spectral capabilities. Thus, radio and X-ray observations,
together with the information on the central black hole mass, should allow
us to study the relationship between the fundamental parameters
characterizing black hole activity such as the central black hole mass,
$M_{\rm BH}$, and the accretion rate, $\dot M$, as well as the
disc--jet coupling.

Until less than ten years ago, quasars and active galactic nuclei, and
the supermassive black holes believed to power them, were regarded as
exceptional (and extreme) objects. However, beginning with the work of
Kormendy \& Richstone (1995) and, in particular, Magorrian et al.
(1998), the idea that SMBH reside in the nuclei of virtually every
galaxy in the nearby universe has become almost commonplace.  
A sharp picture of the
demographics of local SMBH comes from the work of Ho, Filippenko, \&
Sargent (1997b), who have carried out a detailed optical spectroscopic
survey of a large number of nearby galaxies. From this work, it was
found that between a third and a half of the sample has
AGN-like spectra, albeit of low luminosity, (LLAGN; either of the
LINER, Seyfert or Transition type; see Ho, Filippenko, \& Sargent
1997a), thus confirming that SMBH are not only present in galaxies,
but that they are also active (at least as some level). As a matter of
fact, follow-up
radio surveys of optically selected LLAGN \cite{hu01,nag02} have 
yielded extremely high detection rates, with the radio emission having
predominantly a compact core morphology, occasionally accompanied by
jet-like features.  X-ray studies at arcsecond (or sub-arcsecond)
resolution with the {\it Chandra} X-ray Observatory have also allowed
us to firmly pin down the emission properties of the nearby dim
galactic nuclei through systematic surveys of LLAGN
\cite{ho01,ter02,tw03}. These pieces of information on local,
low-luminosity objects nicely complement those on more luminous (and
more distant) ``classical'' AGN (i.e. Seyfert galaxies, QSOs, radio
galaxies) and allow us to investigate the dependences
of observable properties on black hole mass and accretion
rate.

As a matter of fact, 
alongside the work on black hole demographics, there have been various
claims for the existence of a correlation between radio luminosity or
radio loudness (a measure of the ratio of radio to bolometric
luminosity) and SMBH masses \cite{fvf98,mcd01,lac01,nag02,lao00}.
Comparisons of radio emission from GBH and SMBH has also led to the
suggestion that there is a systematic difference in radio loudness
between the two classes, SMBH being on average more radio loud
\cite{fb96}. Furthermore, among X-ray binaries, black holes tend to
be more radio loud than neutron stars \cite{fk01}, possibly also
indicating a mass dependence of the radio loudness parameter. 
However, the difficulty of separating the dependence
of the radio power output on the accretion rate (due to the lack, or
the neglect, of an independent observational indicator of it for SMBH)
unavoidably makes the evidence of any such correlation rather weak.
Indeed, other authors, using different samples, have recently found no
evidence for such correlations \cite{ho02,wu02,oww02}. 

A link between the disc accretion rate and the generation of
relativistic radio jets has been suggested by Willott et al. (1999)
for radio galaxies and by Ho \& Peng
(2001) for Seyfert 1 nuclei, 
on the basis of the strong observed correlations between radio
and optical powers. Moreover, by
studying a sample of galactic nuclei with measured black hole masses,
Ho (2002) found clear evidence of radio loudness being anti-correlated
with some estimator of the dimensionless accretion rate. Interestingly,
Brinkmann et al. (2000) have also correlated the ROSAT All-Sky Survey and
the VLA FIRST catalog, and have shown that there is some degree of
correlation between the monochromatic X-ray luminosity at 2 keV and
the radio (5 GHz) luminosity for bright AGN and Quasars (for both
radio loud and radio quiet sources), while Sambruna et al. (1999) have
found a weak correlation between lobe radio power and 2-10 keV
luminosity in a sample of radio loud AGN.

The situation is rather different for GBH where the measured black
hole masses span a very narrow range (around 10 solar masses). In
this case, the observed, rather large, changes in luminosity can only
be explained if they are somehow linked to variations in the accretion
rate. On this note, recent work has shown that GBHs in the low/hard
state (where their spectra are dominated by a power-law X-ray emission
with typical bolometric luminosities $L_{\rm bol}/L_{\rm Edd} \la
10^{-2}$) always possess compact radio cores \cite{fen01a} 
the radio luminosity of
which is tightly correlated with their X-ray luminosity over more than
three orders of magnitude, with $L_{\rm R}\propto L_{\rm X}^{0.7}$ 
\cite{gfp02,gfp03,cor03}. The evidence of
such a correlation emphasizes the effects of the dependence of the
radio luminosity on the accretion rate 
 and hence the connection between accretion and jet processes.
However, an alternative model has been put forward recently, based on
the broad band spectral energy distributions of some GBH in the
low/hard state, in which the X-ray emission from those objects
is produced by optically thin synchrotron radiation emerging from 
the jet itself \cite{mff01}.  
The tightness of the observed correlation between radio and X-ray
luminosity, and its slope, seem to support such an idea \cite{mar03,cor03}. 

Finally, we should also note that GBH with 
bolometric luminosity close to
their Eddington limit display a more complicated behavior when
observed in the radio band.  There is evidence that in the high/soft
state (when the spectral energy distribution is dominated by a
quasi-thermal component with $kT \sim 1$ keV) continuous jet
production is inhibited, while at still higher luminosities (in the
so-called very high state) powerful, episodic super-luminal ejection
events have been observed (in particular in the prototypical
microquasar GRS 1915+105, Mirabel \& Rodriguez, 1994). Thus, high
luminosity black hole X-ray binaries seem to display a (temporal)
dichotomy between radio loud and radio quiet states,
reminiscent of that of powerful quasars.

This rapid (albeit somewhat disorderly) progress in the observational
studies of black holes of all classes seems to offer the possibility
for understanding fundamental scalings of black hole properties with
$M_{\rm BH}$ and $\dot M$.  Accretion disc theory can provide us with
the relationships between the observed radiative output (hence X-ray
luminosity) and the black hole masses and accretion rates.
Unfortunately, theoretical
modeling of the relation between jet properties (related to the
observed radio luminosity) and the physical parameters of the system
($M_{\rm BH}$, $\dot M$) has been hampered by the lack of a clear
understanding of the mechanism by which jets are accelerated and
collimated. Relating the X-ray and radio properties of a black hole
system to only $M_{\rm BH}$ and/or $\dot M$ has not been straightforward.
However, in a recent paper, Heinz \& Sunyaev (2003) have
demonstrated that, under the general assumption that the jet formation
process is not qualitatively different among SMBH of different mass or
between SMBH and GBH, it is in fact possible to derive a universal
scaling between the jet (radio) luminosity at a given frequency,
$L_{\nu}$, and both mass and accretion rate. The derived
relation is {\it independent of the jet model} and has scaling indices
that depend only on the (observable) spectral slope of the synchrotron
emission in the radio band, and on the accretion model (see
\S~\ref{sec:physics} below).

Motivated by these findings, the aim of this work is twofold.  First,
we want to examine (or re-examine) the significance of the
correlations between radio and X-ray luminosities with $M_{\rm BH}$
and with 
each other in a large sample of black holes with measured masses that
includes {\it both} GBH and SMBH. This will be dealt with in the first
part of the paper: in \S\ref{sec:sam}, we describe the 
selection criteria and the properties of our sample, while in
\S\ref{sec:corr} and \S\ref{sec:results}
we present the results of the correlation analysis and we describe the
observational properties of the fundamental plane of black hole activity.
Then, in the second part of the paper (\S\ref{sec:physics}) 
we will make use of the model
by Heinz \& Sunyaev (2003) to relate observed correlation coefficients
that define the fundamental plane to 
theoretically predicted ones. This will allow us to place constraints
on the physical properties of the accretion flow from which the jet is
launched, and in particular on its radiative efficiency.
Section~\ref{sec:discussion} is devoted to a discussion
of our findings and of their general implications for our
understanding of black hole activity in different regimes. 
Finally, we summarize our results
in \S\ref{sec:summary}

\section{The sample}
\label{sec:sam}
\subsection{Supermassive black holes in galactic nuclei}

We have selected from the existing literature a sample of black
hole-powered systems with measured masses, the nuclei of which have been
observed both at 5 GHz (mostly with arcsecond resolution with the VLA) and
in the 2-10 keV band.  For future reference we define the dimensionless
black hole mass $M=M_{\rm BH}/M_{\odot}$ and accretion rate $\dot m \equiv
(L_{\rm bol}/\eta)/L_{\rm Edd} = \dot M c^2/L_{\rm Edd} \propto \dot M/M$,
where $\eta$ is the accretion efficiency.

We first considered the full sample of $\sim40$ nearby inactive, or weakly
active galaxies with existing nuclear black hole mass measurements from
observations of spatially resolved kinematics. To these we have added a
comparable number of bright AGNs (and QSOs) with nuclear black hole mass
measured from reverberation mapping of their broad line region (a similar
sample was compiled by Ho et al.  2002). From this sample we selected all
objects which have been observed in both the radio and X-ray bands.

In order to obtain a more statistically representative sample, we also
searched the existing literature for both nearby low-luminosity galactic
nuclei (Ho et al., 2001; Terashima et al., 2002; Nagar et al., 2002;
Terashima \& Wilson, 2003) and for relatively bright Seyfert nuclei (either
type 1, type 2 or Narrow Line Seyfert 1) and radio galaxies with available
radio and X-ray flux measurements.  We assign black hole masses to these systems
using the observed correlation between black hole masses and stellar
velocity dispersion \cite{geb00,fm00} \footnote{We use here the
relationship $M = 1.3\times 10^8 (\sigma/200 {\rm km\, s}^{-1})^{4.58}$, as
derived in Ferrarese (2002). See however Tremaine et al. (2002) for a
thorough statistical discussion of the different scalings that have been
claimed for the $M - \sigma$ relation.}, mostly using the values of the
velocity dispersion provided by the HYPERLEDA catalogue \cite{sp02}.
We note here that it is beyond the scope of this paper to discuss the
different methods (and their qualities) 
for estimating black hole masses in the nuclei of
galaxies, and refer the reader to the comprehensive literature on the
subject (see e.g. De Zeeuw, 2003, and references therein).

We do not include in our sample distant quasars, for which the $M-\sigma$ relation cannot be
used because of the lack velocity dispersion measurements. In practice, we
want to avoid any indirect method for black hole mass estimation that
relies on e.g. the continuum optical/UV luminosity (as done by Woo \& Urry,
2002, for example).

Relativistically beamed sources (i.e. those whose jet axis points
towards our line of sight) are dominated by the boosted jet emission,
and cannot be used to test the disc-jet coupling. We therefore
excluded from our sample BL Lac objects. Among the Quasars in our
sample, only 3C 273, which has an extremely high radio loudness and a
blazar-like spectrum, is likely to suffer from strong Doppler boosting
of the radio jet. On the other hand, according to the unification scheme, 
Seyfert 2 nuclei should not be preferentially viewed pole on.
For all the other sources (mainly low-luminosity AGN and Seyfert 1),
for which the nature of the (relatively faint) radio emission is not
well established, we have assumed that the
orientation of their jets with respect to line of sight is randomly
distributed. A more detailed discussion of the possible incidence of
relativistic beaming as a selection effect will be presented in 
\S\ref{sec:beaming}.

\begin{table*}
\caption{Radio and X-ray properties of black holes with measured masses}
\label{tab_1}
\begin{tabular}{lcccccccccc}
\hline
\hline
Object & D(Mpc) & SC & Log $L_{\rm R}$ & $\alpha_{\rm R}$ & Ref & Log $L_{\rm
  X}$ & Ref & Log $M$ & Method & Ref \\
(1) & (2) & (3) &(4) & (5) & (6) & (7) & (8) & (9) & (10) & (11) \\
\hline
Ark 564 & 99.0 & NS1 & 38.59 & U & 1 & 43.74 & 2 & 6.5 & I$_{[OIII]}$
& 3 \\
Cyg A & 224 & S2/L2 & 41.43 & F & 11 & 44.22 & 59 & 9.4 & G & 58\\
Fairall 9  & 199.8 & S1 & $<$37.68 & U & 4 & 44.14 & 5,6 & 7.91 & R & 21 \\
IC 1459 & 29.2 & L2 & 39.71 & F & 4,53 & 40.66$^\dag$ & 9 & 9 & S & 10 \\
IC 4296
 (PKS 1333-33) & 188 & L1.9 & 39.68 & S & 11 & 41.20 & 12 & 9.1
& I$_{\sigma}$ & 13 \\
IC 4329A & 65.5 & S1 & 38.94 & U & 4 & 43.72 & 5 & 6.69 & R & 21 \\
Mrk 3 & 52.0 & S2 & 39.86 & S & 14,51 & 42.7 & 14,15 & 8.81 &
I$_{\sigma}$ & 16 \\
Mrk 78 & 149 & S2 & 39.86 & S & 14,53 & $<$44 & 14 & 7.92 &
I$_{\sigma}$ & 16\\
Mrk 279 & 127 & S1.5 & 38.78 & S & 4,49 & 43.72 & 5 & 7.62 & R & 21 \\
Mrk 335 & 107 & NS1 & 38.27 & F & 4,49 & 43.30 & 2,17 & 6.79 & R & 21\\
Mrk 348 & 60.0 & S2 & 39.70 & F & 14,52 & 42.85 & 14 & 7.17 &
I$_{\sigma}$ & 16\\
Mrk 478 & 316 &NS1 & 38.75 & U & 18 & 43.87 & 2 & 7.30& I$_{\rm [OIII]}$ & 3\\
Mrk 507 & 216 & NS1 &38.78 &U & 1 & 42.97 & 2 & 7.10& I$_{\rm [OIII]}$ & 3\\
Mrk 509 & 144 & NS1 & 38.30$^\S$& S & 19,56 & 44.00 & 20 & 7.86 & R &4\\
Mrk 573 & 69.2 & S2 &38.22 &S & 14,49 & $<$43.00& 14 &  7.25 &
I$_{\sigma}$ & 16 \\
Mrk 590 & 109 & S1.2 &  38.70 & S& 4,49& 43.58 & 57
& 7.23       & R & 21\\
Mrk 766 & 51.7& NS1 & 38.32 & S & 1,22 & 42.90& 2,6 & 6.64 & I$_{\rm [OIII]}$ & 3\\
Mrk 1066 & 48.4 & S2 & 38.68 & S & 14,53 & $<$41.85 & 14 &6.93&
I$_{\sigma}$ &16\\
NGC 221 (M 32)& 0.810& - & $<$33.3& U& 67& 35.97& 67& 6.40& S& 68\\
NGC 315& 66.2 & L1.9 & 40.41& F & 23 & 41.68 & 24 &  9.10&
I$_{\sigma}$ &16\\
NGC 1052 & 19.6 & L1.9 &39.86& S &25& 41.57& 24& 8.29& I$_{\sigma}$ &16\\
NGC 1068 & 14.4 & S1.9& 39.12& S& 4,27&  41.00 & 15& 7.20& M& 4\\
NGC 1275 & 70.1 & S2  & 41.74& F& 14,27& 43.40& 32,34 & 8.64&I$_{\sigma}$
&16\\
NGC 1365 & 21.8 & S1.8& 38.80& S& 53& 40.60 & 24 & 7.66& I$_{\sigma}$&
26\\
NGC 1386 & 11.6& S2& 36.70& U & 53 & 40.64& 24 & 7.20& I$_{\sigma}$&
16\\
NGC 1667 & 61.2& S2& 37.34& S &14,27& 40.05& 15 & 7.93& I$_{\sigma}$&
16\\
NGC 2110 & 31.2& S2& 38.99& F& 14,53& 42.60& 14,15& 8.41& I$_{\sigma}$&16\\
NGC 2273 & 28.4& S2& 37.83& S& 27& 41.40& 15,24& 7.27& I$_{\sigma}$&16\\
NGC 2787 & 7.5& L1.9& 37.22& F& 4,29& 38.40& 24,28& 7.59& G& 10\\
NGC 2841 & 12.0& L2&  36.00& F& 29& 38.26 & 28& 8.42 &  I$_{\sigma}$&30\\
NGC 2992 & 30.8& S2&  38.64& U& 14& 42.10& 14,15& 7.75 &I$_{\sigma}$&16\\
NGC 3031 (M81) &3.9& S1.5& 36.82& F& 4,29& 39.90& 24,28& 7.79& S&4\\
NGC 3079 & 20.4& S2& 38.39& F& 27,50&  40.20& 15& 7.65&I$_{\sigma}$&
26\\
NGC 3147 & 40.9& S2& 38.01& F& 27,31& 41.61& 15,31& 8.79& I$_{\sigma}$& 26\\
NGC 3169 & 16.5& L2& 37.19& F& 31,54& 41.35& 31 & 7.91& I$_{\sigma}$& 33\\
NGC 3226 & 15.4&L1.9& 37.20& F& 31,54& 40.74& 31& 8.23& I$_{\sigma}$& 34\\
NGC 3227 & 20.6&S1.5& 37.94& S& 4,27& 41.99& 5& 7.59& R& 21\\
NGC 3362 & 111& S2& 38.47& S& 14,49& $<$43.6& 14& 6.68& I$_{\sigma}$&16\\
NGC 3516 & 38.9& S1& 37.98& S& 4,27& 43.20& 5,6& 7.36& R& 4\\
NGC 3627 & 6.6& S2& 35.45& F& 29& $<$37.6& 28& 7.26& I$_{\sigma}$&30\\
NGC 3675 & 12.8& T2& $<$35.99& U& 29& $<$37.99& 28& 7.11& I$_{\sigma}$&30\\
NGC 3998 & 14.1& L1.9& 37.98& S& 4,53& 41.66& 24& 8.75& S& 4\\
NGC 4051 & 17.0& NS1& 37.30& S& 4,27& 41.50& 5,6& 6.11 & R& 21\\
NGC 4117 & 124& S2&  35.70& S& 14& 39.4$^\ddag$ & 45& 6.74&
I$_{\sigma}$&16\\
NGC 4143 & 17.0& L1.9& 37.16& F& 29,31& 40.03& 31& 8.32& I$_{\sigma}$& 34\\
NGC 4151 & 20.3& S1.5& 38.49& S& 4,27& 42.83& 5,6& 7.17& R& 21\\
NGC 4203 & 14.1& L1.9& 36.79& F& 29,31& 40.23& 28,31& 7.90& I$_{\sigma}$&30\\
NGC 4258 & 7.3& S1.9& 35.94& F& 4,29& 40.52& 35& 7.60& M& 4\\
NGC 4261 (3C 270)& 29.6& L2& 39.21& U& 11& 41.17& 24& 8.72& G& 4\\
NGC 4278 & 9.7& L1.9& 37.91& F& 29,31& 39.96& 31& 9.20& S& 60\\
NGC 4321 & 16.8& T2& $<$36.17& U& 29& $<$38.59& 28& 6.80&I$_{\sigma}$&30\\
NGC 4374 (M84) &18.4& L2& 38.81& F& 4,29& 40.34& 24& 9.20& G&4\\
NGC 4388 & 16.8& S2& 36.95& S& 27& 42.76& 15& 6.80& I$_{\sigma}$&26\\
NGC 4395 & 3.6& S1.5& 35.56& S& 4,27& 39.50& 36& $<$5.04& S& 4\\
NGC 4450 & 16.8& L1.9& 36.53& F& 29& 40.34& 24& 7.30& I$_{\sigma}$&26\\
NGC 4457 & 17.4& L2& $<$35.70& U& 29& 39.97& 24& 6.86& I$_{\sigma}$&26\\
NGC 4472 & 18.8& S2& 36.69& S& 27,29& $<$38.80& 37& 8.80& I$_{\sigma}$& 38\\
NGC 4486 (M87)& 16.1& L2& 39.78& F& 4,29& 40.55& 39& 9.48& G& 4\\
NGC 4494 & 9.7& L2& $<$35.65& U& 29& 38.86& 28& 7.65&I$_{\sigma}$&30\\
\hline
\end{tabular}
\end{table*}

\begin{table*}
\contcaption{}
\begin{tabular}{lcccccccccc}
\hline
\hline
NGC 4501 & 16.8& S2& 36.20& S& 27& 40.28& 14& 7.90& I$_{\sigma}$&26\\
NGC 4548 & 16.8& L2& 36.30& F& 29& 39.79& 31& 7.40& I$_{\sigma}$&26\\
NGC 4565 & 9.7& S1.9& 36.15& F& 29,31& 39.56& 31& 7.70 & I$_{\sigma}$&26\\
NGC 4579 & 16.8& S1.9& 37.65 & F& 29,31& 41.14& 28,31& 7.85& I$_{\sigma}$&26\\
NGC 4594 (M104) & 9.8& L2& 37.84& F& 4,53& 40.70& 15& 9.04& S& 4\\
NGC 4636 & 17.0& L1.9& 36.40& U& 29& $<$38.43& 37& 7.90&I$_{\sigma}$&
38\\
NGC 4725& 12.4& S2& $<$35.90& U& 29& 39.16& 28& 7.49& I$_{\sigma}$&30\\
NGC 4736 & 4.3& L2& 34.80& F& 29& 39.62& 24& 7.30& I$_{\sigma}$&26\\
NGC 5033 & 18.7& S1.5& 36.79& S& 27& 41.00& 24,28& 7.30& I$_{\sigma}$&30\\
NGC 5194 & 7.7& S2& 35.50& S& 27& 39.80& 15& 6.90& I$_{\sigma}$&16\\
NGC 5252 & 92.3& S2& 38.96& S& 14,49& 42.97& 14& 8.12& I$_{\sigma}$&16\\
NGC 5273 & 21.3& S2& 36.22& F& 14,24& $<$42.0& 14& 6.37&  I$_{\sigma}$&16\\
NGC 5347 & 31.2& S2& 37.10& F& 14,55& 40.0$^\ddag$& 45& 6.70& I$_{\sigma}$&16\\
NGC 5548 & 70.2& S1.5& 38.58& S& 4,27& 43.55& 5& 8.03& R& 21\\
NGC 5929 & 33.2& S2& 38.30& S& 14,49& 40.7$^\ddag$& 45& 7.22& I$_{\sigma}$&16\\
NGC 6166 & 116 & S2& 39.95& S& 40& 40.56& 41& 9.19&  I$_{\sigma}$&26\\
NGC 6251 & 94.8& S2 & 40.93& S& 4,53& 42.15& 15& 8.73& G&4\\
NGC 6500 & 40.0& L2& 38.90& F& 31,54& 40.11& 31& 8.28& I$_{\sigma}$&30\\
NGC 7469 & 66.6& S1& 38.38& S& 4,49& 43.31& 42& 6.81& R& 21\\
NGC 7672 & 53.2& S2& 37.25& S& 14,55& 43.37& 14& 6.80&  I$_{\sigma}$&16\\
NGC 7682 & 68.0& S2& 38.88& S& 14,49& $<$43.2& 14& 7.25& I$_{\sigma}$&16\\
NGC 7743 & 24.4& S2& 36.99& S& 27& 39.71& 24& 6.47& I$_{\sigma}$&16\\
PG 0026+129& 627& Q& 40.17& U& 4& 44.44& 43& 7.73& R& 21\\
PG 0052+251& 690& Q& 39.42& U& 4& 44.66& 44& 8.34& R& 21\\
PG 0804+761& 430& Q& 39.41& U& 4& 44.26& 43& 8.28& R& 21\\
PG 0844+349& 268& Q& 38.12& U& 4& 43.29& 43& 7.34& R& 21\\
PG 0953+414& 1118&Q& 40.14& U& 4& 44.50& 43& 8.26& R& 21\\
PG 1211+143& 362& Q& 41.08& U& 4& 43.61& 43& 7.61& R& 21\\
PG 1226+023 (3C273)& 705& Q& 44.03& F& 4,56& 45.70& 44& 8.74& R& 21\\
PG 1229+204 & 268& Q& 38.45& U& 4& 44.29& 44& 7.88& R& 21\\
PG 1307+085 & 690& Q& 38.98& U& 4& 44.51& 44& 8.44& R& 21\\
PG 1411+442 & 380& Q& 38.71& U& 4& 43.53& 43& 7.90& R& 21\\
PG 1426+015 & 366& Q& 38.98& U& 4& 43.89& 44& 8.67& R& 21\\
PG 1613+658 & 565& Q& 39.75& F& 4,56& 44.34& 44& 8.38& R& 21\\
PG 1700+518 & 1406& Q& 40.92& S& 4,53& $<$43.20& 43& 7.78& R& 21\\
PG 2130+099 & 255& Q& 38.89& S& 4,56& 43.55& 44& 8.16& R& 21\\
3C120& 138& S1& 41.55& F& 11,53& 43.95& 11& 7.36& R& 21\\
3C 390.3& 241& S1& 41.09& S& 11& 44.00& 11& 8.53& R& 21\\
UGC 6100& 116& S2& 38.50& U& 14& $<$43.6& 14& 7.72&  I$_{\sigma}$&16\\
Sgr A$^*$& 0.008 &-& 32.50& F& 62 & 33.34& 63& 6.41& S& 64\\
Cyg X-1& 0.0021& GBH& $<$29.44& F& 61& 36.44& 61& 1.06& S& 69\\
Cyg X-1& 0.0021& GBH& $<$29.45& F& 61& 36.57& 61& 1.06& S& 69\\
Cyg X-1& 0.0021& GBH& 29.63& F& 61& 36.48& 61& 1.06& S& 69\\
Cyg X-1& 0.0021& GBH& 29.65& F& 61& 36.57& 61& 1.06& S & 69\\
Cyg X-1& 0.0021& GBH& 29.66& F& 61& 36.64& 61& 1.06& S & 69\\
Cyg X-1& 0.0021& GBH& 29.24& F& 72& 36.58& 61& 1.06& S & 69\\
Cyg X-1& 0.0021& GBH& 29.30& F& 72& 36.65& 61& 1.06& S& 69\\
Cyg X-1& 0.0021& GBH& 29.18& F& 72& 36.66& 61& 1.06& S& 69\\
Cyg X-3& 0.009& GBH& $<$30.64& F& 61& 36.94& 61& 1.00& S& 70\\
Cyg X-3& 0.009& GBH& 31.62& F& 61& 37.19& 61& 1.00& S& 70\\
Cyg X-3& 0.009& GBH& 31.85& F& 61& 37.41& 61& 1.00& S& 70\\
Cyg X-3& 0.009& GBH& 32.17& F& 61& 37.60& 61& 1.00& S& 70\\
Cyg X-3& 0.009& GBH& 32.43& U& 61& 37.88& 61& 1.0& S& 70\\
GRO J1655-40& 0.0032& GBH& $<$29.76& F& 61& 36.64& 61& 0.85& S& 65\\
GRO J1655-40& 0.0032& GBH& $<$29.69& F& 61& 37.55& 61& 0.85& S& 65\\
GRO J1655-40& 0.0032& GBH& 29.94& F& 61& 35.27& 61& 0.85& S& 65\\
GRS 1915+105& 0.0125& GBH& $<$30.89& F& 61& 38.17& 61& 1.20& S& 65\\
GRS 1915+105& 0.0125& GBH& $<$30.89& F& 61& 38.47& 61& 1.20& S& 65\\
GRS 1915+105& 0.0125& GBH& $<$30.89& F& 61& 38.73& 61& 1.20& S& 66\\      
GRS 1915+105& 0.0125& GBH& 31.60& F& 61& 38.25& 61& 1.20& S& 66\\
GRS 1915+105& 0.0125& GBH& 31.76& F& 61& 38.40& 61& 1.20& S& 66\\
GRS 1915+105& 0.0125& GBH& 31.76& F& 61& 38.65& 61& 1.20& S& 66\\
\hline
\end{tabular}
\end{table*}

\begin{table*}
\contcaption{}
\begin{tabular}{lcccccccccc}
\hline
\hline
GX 339-4& 0.004& GBH& 29.91& F& 7& 36.48& 7& 1.00& S& 8\\
GX 339-4& 0.004& GBH& 29.87& F& 7& 36.42& 7& 1.00& S& 8\\
GX 339-4& 0.004& GBH& 29.89& F& 7& 36.40& 7& 1.00& S& 8\\
GX 339-4& 0.004& GBH& 29.62& F& 7& 36.12& 7& 1.00& S& 8\\
GX 339-4& 0.004& GBH& 29.71& F& 7& 36.12& 7& 1.00& S& 8\\
GX 339-4& 0.004& GBH& 29.66& F& 7& 36.14& 7& 1.00& S& 8\\
GX 339-4& 0.004& GBH& 29.45& F& 7& 35.81& 7& 1.00& S& 8\\
GX 339-4& 0.004& GBH& 29.11& F& 7& 35.32& 7& 1.00& S& 8\\
GX 339-4& 0.004& GBH& 28.34& F& 7& 34.21& 7& 1.00& S& 8\\
GX 339-4& 0.004& GBH& 28.02& F& 7& 33.91& 7& 1.00& S& 8\\
GX 339-4& 0.004& GBH& 28.38& F& 7& 33.84& 7& 1.00& S& 8\\
GX 339-4& 0.004& GBH& 28.49& F& 7& 34.01& 7& 1.00& S& 8\\
GX 339-4& 0.004& GBH& $<$27.51& F& 7& $<$33.34& 7& 1.00& S& 8\\
GX 339-4& 0.004& GBH& $<$27.25& F& 7& 33.19& 7& 1.00& S& 8\\
LS 5039 & 0.003& GBH& 29.80& S& 61& 35.62& 61& 1.00&- &- \\
LS 5039 & 0.003& GBH& 30.09& S& 61& 35.45& 61& 1.00&- &- \\
LS 5039 & 0.003& GBH& 30.07& S& 61& 35.67& 61& 1.00&- &- \\
LS 5039 & 0.003& GBH& 30.09& S& 61& 35.81& 61& 1.00&- &- \\
LS 5039 & 0.003& GBH& 30.08& S& 61& 35.96& 61& 1.00&- &- \\
XTE J1118+480& 0.0018& GBH& $<$28.98& F& 46& 35.43& 47& 1.00& S& 48 \\
XTE J1118+480& 0.0018& GBH& 28.92& F& 71& 35.46& 61& 1.00& S& 48\\
XTE J1118+480& 0.0018& GBH& 28.92& F& 71& 35.57& 61& 1.00& S& 48\\
XTE J1118+480& 0.0018& GBH& 28.92& F& 71& 35.56& 61& 1.00& S& 48\\
XTE J1118+480& 0.0018& GBH& 28.92& F& 71& 35.47& 61& 1.00& S& 48\\
XTE J1118+480& 0.0018& GBH& 28.92& F& 71& 35.45& 61& 1.00& S& 48\\
XTE J1859+226& 0.011& GBH& $<$29.24& F& 61& 36.58& 61& $>$0.88& S& 71\\
XTE J1859+226& 0.011& GBH& $<$29.30& F& 61& 36.65& 61& $>$0.88& S& 71\\
XTE J1859+226& 0.011& GBH& 29.18& F& 61& 36.66& 61& $>$0.88& S& 71\\
\hline
\end{tabular}
\vskip 0.3cm NOTE: Comments: $^\dag$ $L_{\rm X}$ calculated from the known
luminosity in the 0.3-8 keV band and the observed spectral index; $^\ddag$
$L_{\rm X}$ calculated from the GIS count rate, assuming $\Gamma=2$ and
$Log (N_{\rm H})=22.5$; $^\S$ $L_{\rm\,5Ghz}$ extrapolated from observations
at 8 GHz. Col.(1): Name of the object. Col. (2): Distance in Megaparsecs
(for $H_0=75$ km s$^{-1}$ Mpc$^{-1}$). Col. (3): Spectral Class; GBH:
Galactic black hole; L: LINER; S: Seyfert; NS1: Narrow Line Seyfert 1; T:
Transition object (LINER/H II); Q: Quasar. Col. (4) Logarithm of nuclear
luminosity at 5GHz. Col. (5): Radio spectral index $\alpha_{\rm R}$ (where
$F_{\nu}=\nu^{-\alpha_{\rm R}}$); F: flat spectrum ($\alpha_{\rm
  R}<0.4$); S: steep 
spectrum ($\alpha_{\rm R}>0.4$); U: undetermined. Col. (7) Logarithm of the
intrinsic rest-frame luminosity in the 2-10 keV band. Col. (9) Logarithm of
the black hole mass. Col. (10) Mass measurement method; S: stellar
kinematics; G: gas kinematics; M: maser kinematics; R: reverberation
mapping; I$_{\rm [OIII]}$: inferred from the mass-[OIII] line-width
correlation; I$_{\sigma}$: inferred from the mass-velocity dispersion
correlation.

REFERENCES: (1) Ulvestad, Antonucci \& Goodrich (1995); (2) Leighly (1999);
(3) Wang \& Lu (2001); (4) Ho (2002); (5) Weaver, Gelbord \& Yaqoob (2001);
(6) Nandra et al. (1997); (7) Corbel et al. (2003); (8) Hynes et
al. (2003); (9) Fabbiano et al. (2003); (10) Tremaine et al. (2002); (11)
Sambruna, Eracleous \& Mushotzky (1999); (12) Pellegrini et al. (2003);
(13) Saglia et al. (1993); (14) Polletta et al. (1996); (15) Bassani et
al. (1999); (16) Woo \& Urry (2002); (17) Gondoin et al. (2002); (18) Zhou
\& Wang (2002); (19) Thean et al. (2001); (20) Pounds et al. (2001); (21)
Kaspi et al. (2000); (22) Ho \& Peng (2001); (23) Gregory \& Condon (1991);
(24) Terashima et al. (2002); (25) Griffith et al. (1995); (26) Simien \&
Prugniel (2002) and HYPERLEDA Catalogue:
http://www-obs.univ-lyon1.fr/hypercat/; (27) Ho \& Ulvestad (2001); (28) Ho
et al. (2001); (29) Nagar et al. (2002); (30) Barth, Ho \& Sargent (2002);
(31) Terashima \& Wilson (2003); (32) Allen et al. (2001); 
(33) Hereudeau \& Simien (1998); (34) Churazov et al. (2003); (35)
Fabbiano et al. (1992); (36) Shih et al. (2003); (37) Loewenstein et
al. (2001); (38) Merritt \& Ferrarese (2001); (39) Di Matteo et al. (2003);
(40) Giovannini et al. (1998); (41) Di Matteo et al. (2001); (42) De Rosa,
Fabian \& Piro (2002); (43) George et al. (2000); (44) Lawson \& Turner
(1997); (45) Moran et al. (2001); (46) Dhawan et al.  (2000); (47) Frontera
et al. (2001); (48) Wagner et al. (2001); (49) Rush, Malkan \& Edelson
(1996); (50) Becker, White \& Edwards (1991); (51) Kojoan et al. (1980);
(52) White, Giommi \& Angelini (2000); (53) V\'eron-Cetty \& V\'eron
(2001); (54) Falcke et al. (2001); (55) Ulvestad \& Wilson (1989); (56)
Falcke, Malkan \& Biermann (1995); (57) Turner \& Pounds (1989); (58)
Tadhunter et al. (2003); (59) Young et al. (2002); (60) Magorrian et
al. (1998); (61) This work; (62) Melia \& Falcke (2001); (63) Baganoff et
al. (2001); (64) Sch\"odel et al. (2002); (65) Orosz \& Bailyn (1997); (66)
Greiner, Cuby \& McCaughrean (2001); (67) Ho, Terashima \& Ulvestad (2003);
(68) Verolme et al. (2002); (69) Dolan (1992); (70) Hanson, Still \& Fender
(2000); (71) Filippenko \& Chornock (2001); (72) Stirling et al. (2001),
(73) Fender et al. (2001); .
\end{table*}

\subsection{Galactic black hole sources}
\label{sec:sam_gbh}
The Galactic X-ray binaries included in our sample have been selected
to have (a) simultaneous X-ray and radio observations, or {\em RXTE}
All-Sky-Monitor (ASM) X-ray data in conjunction with radio fluxes
available from the literature, and (b) publicly available {\em
RXTE}-ASM X-ray and Green-Bank Interferometer (GBI) radio lightcurves
(from which we estimated the 5 GHz fluxes by interpolating between the
2.25 GHz and the 8.3 GHz channels).  

We treated the GBI and ASM data of each object in the following way: we
first re-gridded the ASM X-ray lightcurve to the radio lightcurve, and then
sorted the data into bins of ascending X-ray flux.  This procedure assumes
that individual bins are not correlated and can thus be re-ordered and
rebinned to improve statistics. Rebinning the data in time intervals does
not change the results significantly, but reduces the dynamic range in
X-ray flux. In this sense, X-ray-flux-ordered binning produces a better
representation of the scatter in this variable. Since the GBI becomes noise
dominated below about 15\,{mJy}, we conservatively chose to consider data
at or below this value as upper limits and split the sample into detection
and upper limits before binning. ASM fluxes were converted to $2-10\,{\rm
keV}$ luminosities using the a conversion factor of $3.2\times
10^{-10}\,{[\rm erg\,s^{-1}\,cm^{-2}]/[cts\,s^{-1}]}$ \cite{ggs02}, which
assumes a crab-like source spectrum.

Black hole masses for GBH are derived from stellar kinematics and we
use the value provided in the literature.  For the microquasar 
LS 5039, no mass estimate is available, and we assumed $M_{\rm BH} =
10\,M_{\odot}$. 

\begin{figure*}
\begin{tabular}{cc}
\psfig{figure=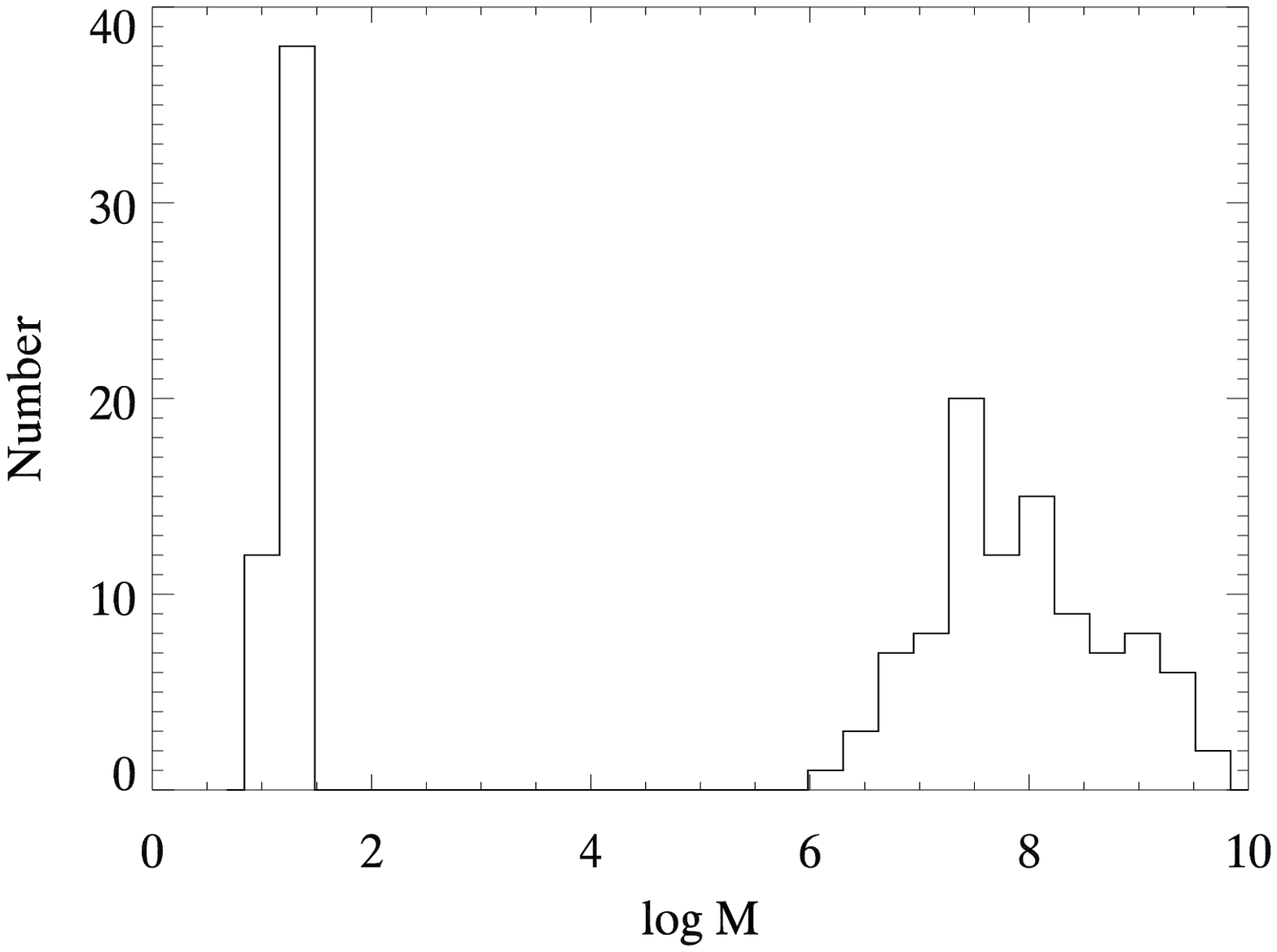,angle=0,width=0.49\textwidth}
&\psfig{figure=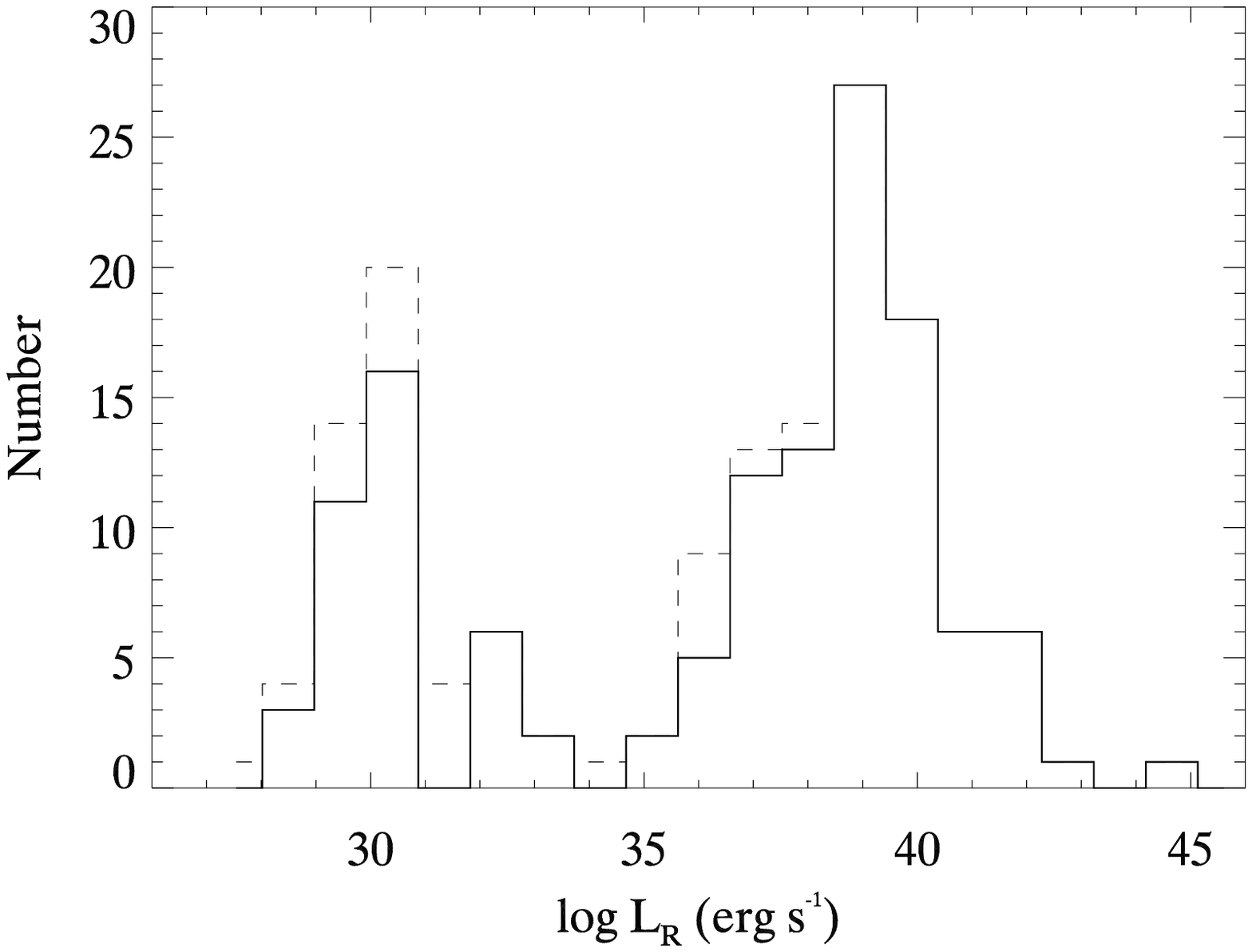,angle=0,width=0.49\textwidth}\\
\psfig{figure=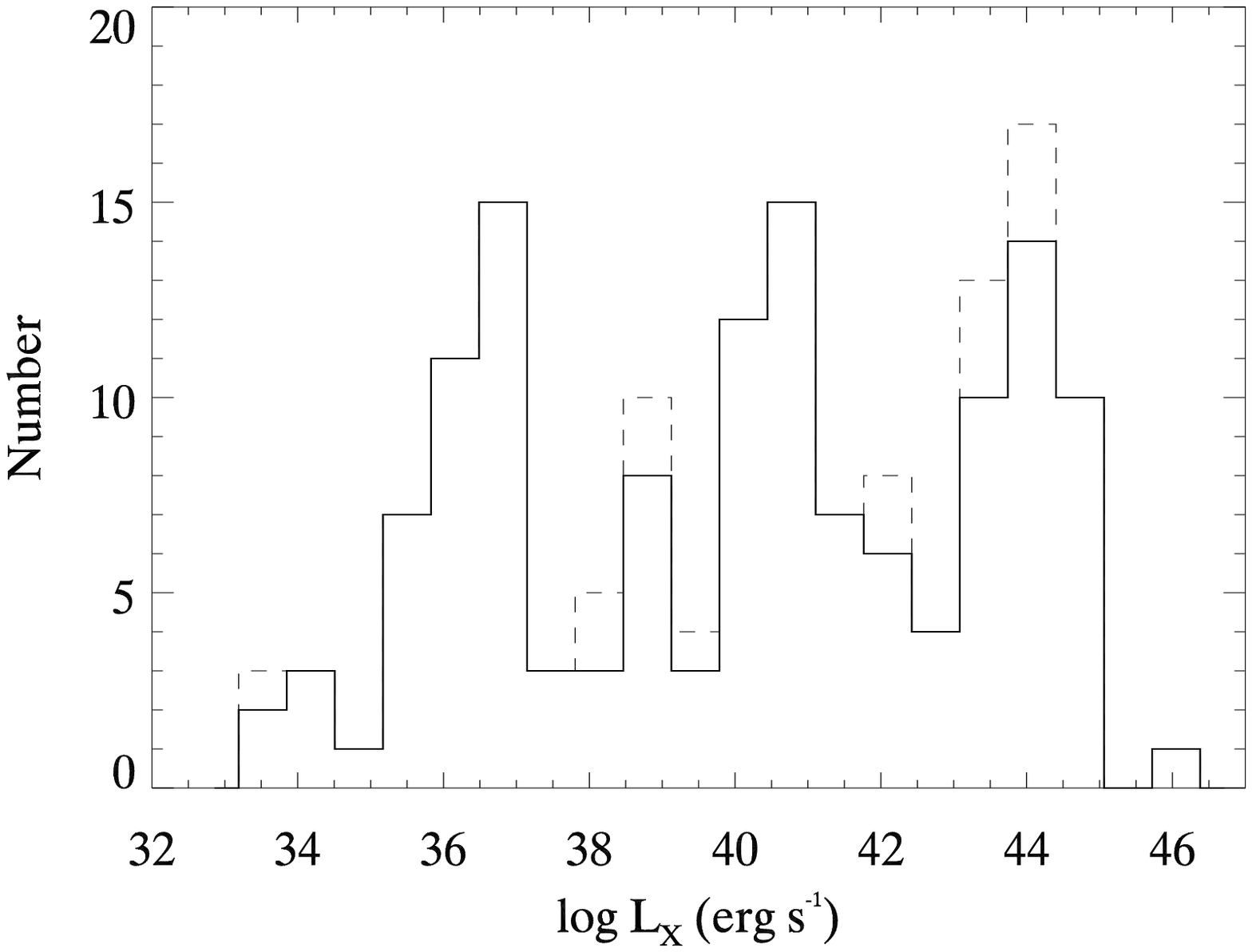,angle=0,width=0.49\textwidth}
&\psfig{figure=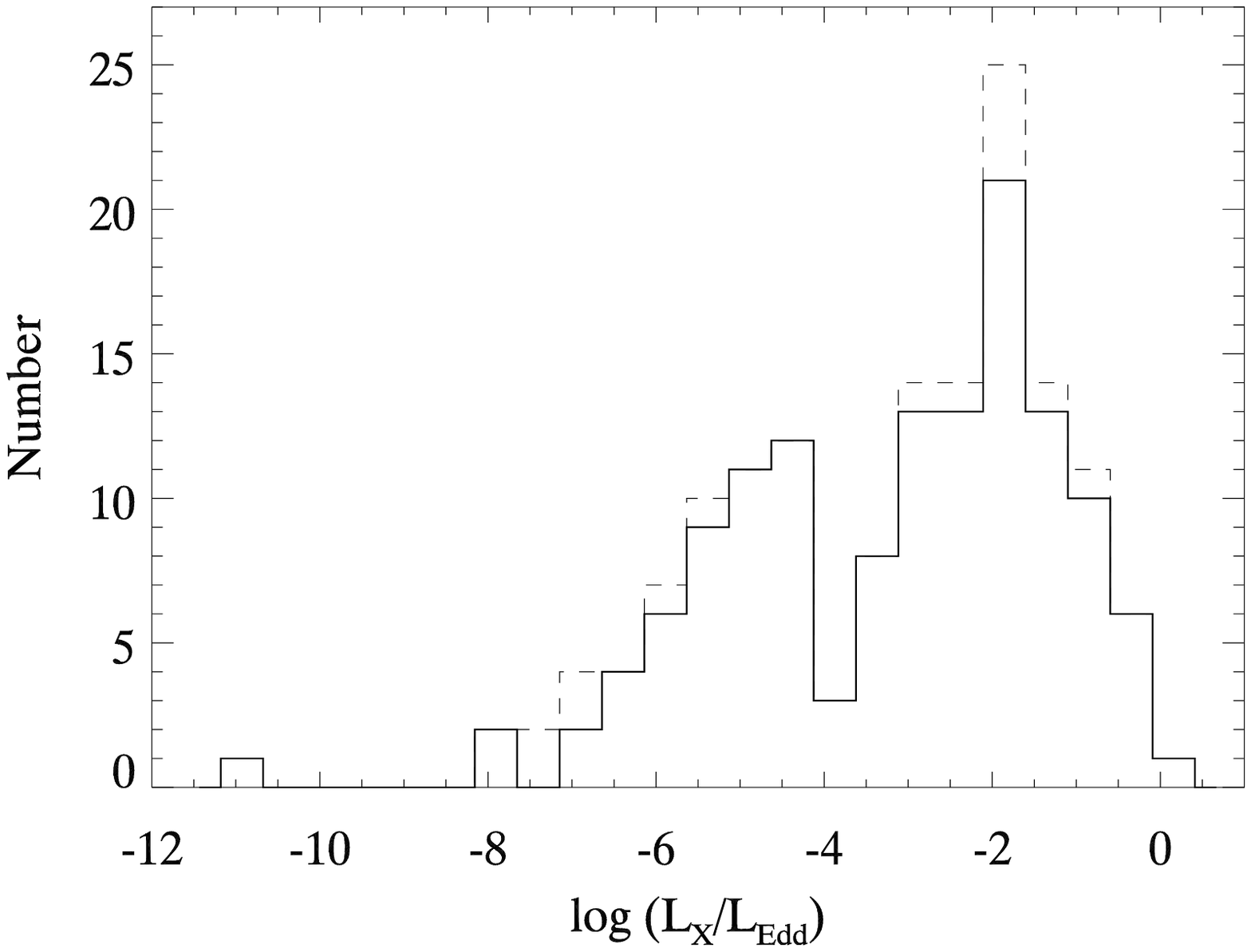,angle=0,width=0.49\textwidth}\\
\end{tabular}
\vskip 0.4cm
\caption{Distribution of black hole masses (top left panel), 5 GHz
  radio luminosity ($L_{\rm R}$, top right panel), 2-10 keV X-ray
  luminosity ($L_{\rm X}$, bottom right panel) and of the ratio
  $L_{\rm X}/L_{\rm Edd}$. Solid histograms denote detected sources,
  dashed ones are for upper limits.} 
\label{fig:histo}
\end{figure*}

\subsection{Global properties of the sample}
Our final sample (Table~\ref{tab_1}) comprises 149 entries, of which 99 are
individual SMBH systems.  The remaining 50 correspond to the observations
of 8 different GBH at various luminosity levels. Out of these 8 sources, 5
are so-called X-ray transients (Cyg X-3, GRO J1655-40, GRS 1915+105, XTE
J1118+480 and XTE J1859+226) and 3 persistent sources (Cyg X-1, GX 339-4
and LS 5039). The SMBH sample includes 14 Quasars; 19 Seyfert galaxies of
type 1 (in this class we include all the Seyfert spectrally classified as
types 1 to 1.9); 32 Seyfert 2 galaxies; 7 Narrow Line Seyfert 1 galaxies
(NS1); 11 Low-Ionization Nuclear Emission line Regions (LINERS) of type
1.9; 13 LINERS of type 2 and 2 LINER/HII Transition nuclei. Separating mass
estimators into direct ones (stellar kinematics, maser emission, gas
kinematics and, to a lesser extent, reverberation mapping) and indirect
ones (all those that infer a measure of the central black hole mass from
the observed $M-\sigma$ relation), we can assess possible biases introduced
in our sample by the different mass measurement techniques. 55 masses are
measured indirectly (4 Sy1, 4 NS1, 30 Sy2, 15 LINERs and the two
transition objects) and 44 directly (all the 14 QSOs, 15 Sy1, 2 Sy2, 3
NS1, 8 LINERs plus Sgr A* and M 32).

For the radio and X-ray luminosities taken from the literature, we have
assumed $H_0 = 75$ km s$^{-1}$ Mpc$^{-1}$, correcting the quoted values
when necessary.  When available, we have included information on the radio
spectral properties of the sources. In Table~\ref{tab_1} we have marked all
the sources with flat radio spectrum ($\alpha_{\rm R}<0.4$, where
$\alpha_{\rm R}$ is the radio spectral index $\alpha \equiv - \partial
\ln{L_{\nu}}/\partial \ln{\nu}$ evaluated at the frequency $\nu=$5 GHz)
with flag (F). All the sources with $\alpha_{\rm R}>0.4$ are instead
classified as steep spectrum (S), while those for which the radio spectral
index couldn't be determined are marked with a (U).  Overall, we have 77
flat spectrum sources (32 SMBH and 45 GBH, respectively); 43 steep spectrum
sources (38 and 5) and 29 for which the spectral slope is undetermined (all
SMBH).

In Figure~\ref{fig:histo} we show the distributions of black hole
masses, radio and X-ray luminosities and that of the ratio of their
X-ray to the Eddington luminosity, $L_{\rm X}/L_{\rm Edd}=L_{\rm 2-10
  keV}/1.3 \times 10^{38} M$.  It is worth emphasizing that the shape
of those distributions reflects more the nature of our selection
procedure rather than the global intrinsic properties of accreting
black holes. For example, a precise measurement of the central black
hole mass is a prerequisite for a given source to be included in the
sample. As recently discussed by De Zeeuw (2003), not all mass ranges
are equally well probed by the different methods, and this should
introduce a strong selection effect in our sample. Similarly, because
BH masses are more easily measured in the nuclei of nearby galaxies,
our sample is biased against the most luminous quasars. This should be
taken into account when examining, for example, the $L_{\rm X}/L_{\rm
  Edd}$ distribution of the black holes in our sample.

In Figure~\ref{fig_lr_m_class} we show the radio luminosity versus the
black hole mass for objects of different spectral classes. Panel (a)
shows the whole sample, while panel (b) concentrates on the SMBH
only. Overplotted are some of the linear regression fits discussed in
the recent literature (see \S\ref{sec:corr} for details).

Finally, Figure~\ref{fig:lr_lx_bin} shows the core radio luminosity versus
the nuclear X-ray luminosity in the 2-10 keV band (left panel) and versus
the ratio of the X-ray nuclear luminosity to the Eddington luminosity
(right panel).  We represent objects in different mass bins with different
colors to highlight a possible segregation of different mass bins in the
$L_{\rm R}-L_{\rm X}$ plane.

In the next section we present a more quantitative statistical analysis
of the correlations among $L_{\rm R}$, $L_{\rm X}$, and $M$.

\section{Correlation analysis}
\label{sec:corr}
Luminosity-luminosity correlations have to be tested for possible
spurious effects introduced by their common dependence on the
distance. Such tests can be done by performing a partial correlation
analysis, taking distance as the third variable.  Here we choose the
so-called partial Kendall's $\tau$ correlation test, proposed by
Akritas \& Seibert (1996) in the case of censored data sets.
Applying this test to our data, we find that the $L_{\rm R} - L_{\rm X}$
correlation is strong: the probability of the null hypothesis (i.e. that
there is no correlation) is less that $10^{-10}$ for the whole sample and
$\sim 1.12 \times 10^{-4}$ for the SMBH only (see Table~\ref{tab:corr}).

\begin{figure*}
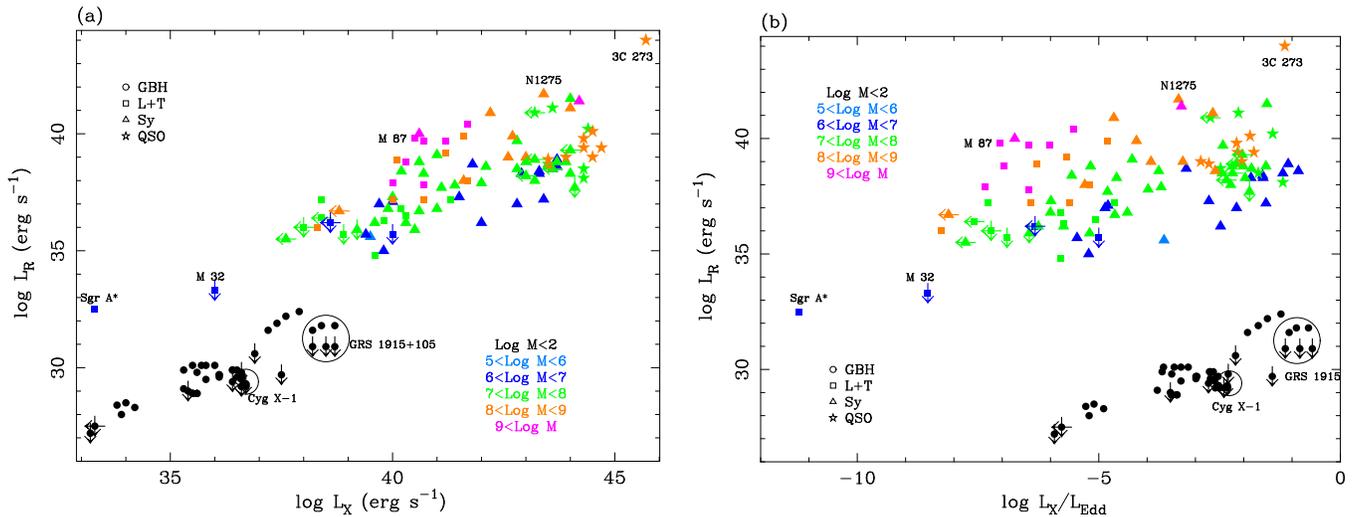

\begin{tabular}{cc}
\psfig{figure=lr_lx_bin.ps,angle=270,width=0.49\textwidth}
&\psfig{figure=lr_mdotx_bin.ps,angle=270,width=0.49\textwidth}\\
\end{tabular}
\caption{Radio core luminosity at 5 GHz versus X-ray luminosity in the
  2-10 keV band (a) for the whole sample. Different colors correspond
  to different mass bins, while different symbols correspond to
  different classes of objects. In panel (b) we plot the same radio
  core luminosity at 5GHz vs. the ratio $L_{\rm X}/L_{\rm Edd}$ of
  X-ray to Eddington luminosity. The color-coding of the different mass
  bins makes the mass segregation more evident.} \label{fig_lr_lx_bin}
\label{fig:lr_lx_bin}
\end{figure*}

However, it is apparent from Fig.~\ref{fig_lr_lx_bin} that, when the
data points are grouped into mass bins, objects in different bins tend
to lie on parallel tracks. Such a behavior can be seen both in the
$L_{\rm R}-L_{\rm X}$ plane (panel a), and, perhaps even more clearly,
in the $L_{\rm R}-L_{\rm X}/M$ plane (panel b). The presence of a mass
segregation suggests that the radio luminosity of an object likely
depends both on its X-ray luminosity and on its mass. In order to
assess this hypothesis we once again use partial correlation analysis,
now taking $L_{\rm R}$ (or $L_{\rm X}$) as the dependent variable and
testing its partial correlation with $L_{\rm X}$ ($L_{\rm R}$) where
we take $M$ as the third variable. In addition, we also look for a
partial correlation between $L_{\rm R}$ ($L_{\rm X}$) with mass
itself, by taking $L_{\rm X}$ ($L_{\rm R}$) as the third variable. In
so doing we can effectively discriminate between intrinsic and
spurious correlations among the three quantities.

The results of such tests (see Table~\ref{tab:corr}) show that the radio
luminosity is strongly correlated with both black hole mass and X-ray
luminosity ($P_{\rm null}< 1 \times 10^{-10}$ for the whole data set),
while in turn, the X-ray luminosity correlates with both mass and radio
luminosity only if we include both GBH and SMBH 
($P_{\rm null} \simeq 4.68 \times 10^{-5}$). 
These results imply that any regression fit used to find correlations
between any two variables that does not account for the dependence on the
third one (as, for example trying to find the dependence of $L_{\rm R}$ on
$M$ without accounting for the dependence on $L_{\rm X}$), inevitably leads
to an incorrect estimate of the correlation coefficients.

\begin{figure*}
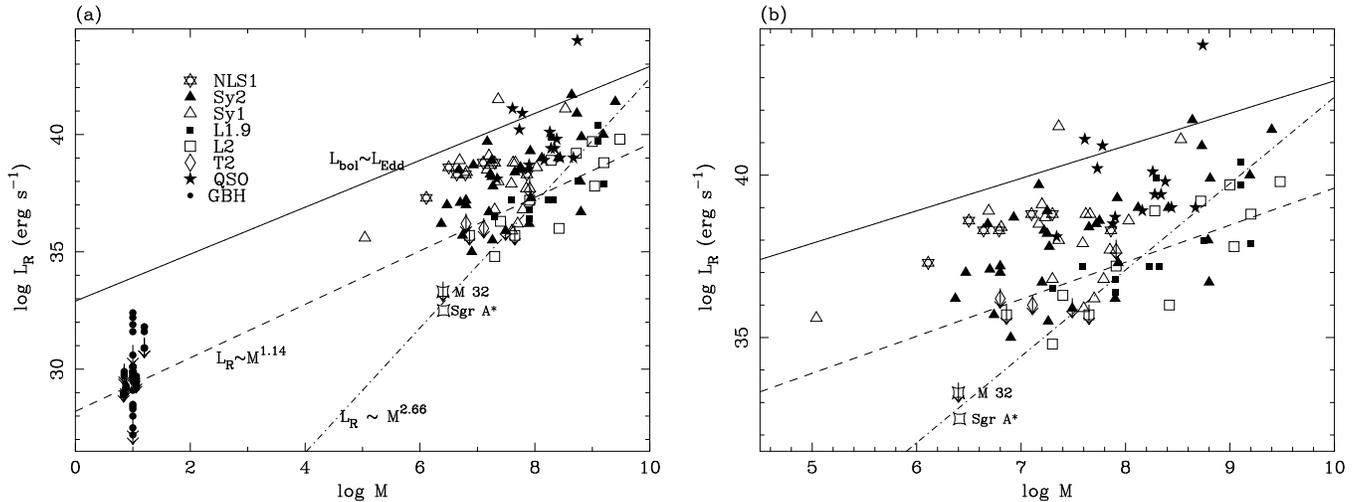

\begin{tabular}{cc} \psfig{figure=lr_m_class_bw.ps,angle=270,width=0.49\textwidth}
& \psfig{figure=lr_m_class_smbh_bw.ps,angle=270,width=0.49\textwidth}\\
\end{tabular}
\caption{Radio core luminosity at 5 GHz versus black hole mass. Upper
  limits are marked with arrows. 
  Panel (a) shows the whole sample, including GBH (black filled
  circles) and SMBH, with different symbols indicating objects
  belonging to different spectral classes. The right panel (b) is a
  blow up of the SMBH sample. The dot-dashed line gives the regression
  fit proposed by Franceschini et al. (1998), the  dashed line
  that proposed by Nagar et al. (2002), both obtained using different
  samples of SMBH only. The thick solid upper line gives the maximum
  core radio power as calculated by Ho (2001) for sources accreting at
  the Eddington rate. Although a correlation analysis of our sample
  would give results similar to those obtained by Nagar et al. (2002),
  as we discuss in the text, {\it none} of the above relations
  reflects the real physical scaling of radio power and black hole mass.} 
\label{fig_lr_m_class}
\end{figure*}

If, in any case, we compute the correlation coefficients between any
two of these variables\footnote{We use a the linear regression method
by parametric EM algorithm (that deals with censored data) as
implemented in the ASURV package \cite{ifn86}} to allow a comparison  with
earlier works, we find that $\log L_{\rm R}^{\rm SMBH} = (29.54\pm
1.60)+(1.23 \pm 0.20)\log M$, with a very large scatter (standard
deviation $\sigma_{\rm RM}^{\rm SMBH}= 1.65$). Including GBH, we
find $\log L_{\rm R}^{\rm all} = (28.75\pm 0.18)+(1.20 \pm 0.04)\log
M$ (standard deviation $\sigma_{\rm RM}^{\rm all}= 1.51$), which is
indeed very similar to what found by Nagar et al. (2002), but very
different from what was proposed by Franceschini et al. (1998), based
on a much smaller sample (as shown by the different linear regression
fits plotted in Figure~\ref{fig_lr_m_class}).

We emphasize, however, that the partial correlation analysis presented
so far and summarized in Table~\ref{tab:corr}, implies that a far better
representation of the mutual dependencies of $L_{\rm R}$, $L_{\rm X}$
and $M$ (and the one that minimizes the scatter) 
should be searched for with multivariate linear regression
tests, that allow simultaneous fitting over the 3-dimensional space defined by
the three variables.

\begin{table*}
\caption{Results of Correlation analysis}
\label{tab:corr}
\begin{tabular}{llllccccccc}
\hline
\hline
\multicolumn{3}{c}{Variables} &  \multicolumn{5}{c}{Subsample} &
\multicolumn {3}{c}{Correlation}\\
\multicolumn{3}{c}{} &  \multicolumn{5}{c}{} & \multicolumn {3}{c}{}\\
X & Y & Z & Objects & N& N$^X_{\rm ul}$ & N$^Y_{\rm ul}$ &  N$^Z_{\rm
  ul}$& $\tau$& $\sigma$ & P${\rm null}$ \\
(1) & (2) &(3) &(4) & (5) &(6) & (7) &(8) &(9) & (10) &(11)\\
\hline
Log $L_{\rm R}$ & Log $L_{\rm X}$ & Log $D$ & All Objects & 149 & 20 & 14
& 0 & 0.255 & 0.0414 &  $7.3 \times 10^{-10}$\\
Log $L_{\rm R}$ & Log $L_{\rm X}$ & Log $D$ & SMBH only & 99 & 7 & 13
& 0 & 0.200 & 0.0518 & 1.12 $\times 10^{-4}$ \\
Log $L_{\rm R}$ & Log $L_{\rm X}$ & $M$ & All Objects & 149 & 20 & 14
& 1 & 0.448 & 0.0495 &   $<1\times 10^{-10}$\\
Log $L_{\rm R}$ & Log $L_{\rm X}$ & $M$ & SMBH only & 99 & 7 & 13
& 1 & 0.450 & 0.0523 & $<1\times 10^{-10}$ \\
Log $L_{\rm R}$ & $M$& Log $L_{\rm X}$ & All Objects & 149 & 20 & 1
& 14 & 0.432 & 0.0469 &  $<1\times 10^{-10}$\\
Log $L_{\rm R}$ & $M$ & Log $L_{\rm X}$ & SMBH only & 99 & 7 & 1
& 13 & 0.310 & 0.0547 & 1.45 $\times 10^{-8}$\\
Log $L_{\rm X}$ & $M$& Log $L_{\rm R}$ & All Objects & 149 & 14 & 1
& 20 & 0.184 & 0.0452 & $4.68 \times 10^{-5} $\\
Log $L_{\rm X}$ & $M$ & Log $L_{\rm R}$ & SMBH only & 99 & 13 & 1
&7 & -0.022 & 0.052 & 0.672\\
\hline
\end{tabular}
\vskip 0.3cm NOTE: Col. (1): Variable X. Col. (2): Variable Y. Col (3):
Variable Z. Correlation between variables X and Y is studied, taking into
account the mutual correlation of X nd Y with Z. Col. (4):
subsample. Col. (5): Number of objects in the subsample. Col. (6)-(8):
Number of upper limits in X, Y and Z. Col. (9)-(11): Results of partial
correlation analysis, giving the partial Kendall's $\tau$ correlation
coefficient, the square root of the calculated variance $\sigma$, and the
associated probability $P_{\rm null}$ for accepting the null hypothesis
that there is no correlation between X and Y.
\end{table*}

\subsection{Finding the multivariate correlation coefficients} 
The standard multivariate regression formalism does not treat dependent and
independent variables symmetrically (see, e.g., Fasano \& Vio 1988).  In other
words, if a linear regression analysis yields a linear regression
coefficient $b$ for the dependence of $y$ on $x$, it does not necessarily
yield a coefficient $b^{-1}$ for the dependence of $x$ on $y$.  Closely
related to this is the fact that standard $\chi^2$ statistics, for which
errors are only associated with the dependent variable, artificially reduce
the estimate of the correlation coefficient in the presence of intrinsic
scatter in the independent variable. The net result of these shortcomings
is that $\chi^2$ distributions over the fitting parameter space are not
fair representations of the confidence in the resulting best fit.

In order to alleviate these shortcomings, we extend the statistical
approach used by the ``Nukers'' group \cite{geb00,tre02} to obtain an
unbiased estimator of the best fit regression coefficients in multivariate
problems. This approach uses a modified chisquare estimator, called the
merit function \cite{fv88,pre92,tre02}, defined by
\begin{equation}
        \hat{\chi}^2=\sum_{i}\frac{(y_{i} - a - \sum_{j} b_{
        j} x_{ij})^2}{\sigma_{y_{i}}^2 + \sum_{j} (b_{
        j}\sigma_{x_{ij}})^2} \label{eq:merit}
\end{equation}
where $y_{i}$ is the dependent variable, $x_{ij}$ are the
independent variables and $\sigma_{y_{i}}$ and $\sigma_{x_{ij}}$
the associated error estimates, $a$ is the zero intercept, and ${\mathbf
b}$ are the linear regression coefficients, which are to be found.  The
argument of the sum in eq.~(\ref{eq:merit}) is a measure of the projected
distance of the data point $P_{i}=(x_{ij},y_{i})$ to the
regression hyperplane given by the equation $y=a+\sum_{j} b_{
j}x_{j}$, measured in the $\chi^2$ space around point $P_{i}$.  In
other words, the argument of the sum in eq.~(\ref{eq:merit}) counts the
number of $\chi^2$ intervals between $P_{i}$ and the regression
hyperplane.  Because eq.~(\ref{eq:merit}) is nonlinear in $b_{j}$, it
is no longer possible to minimize $\hat{\chi}^2$ analytically.  However,
for a set of parameters ${\mathbf b}$, we can still find the optimal value
$a_{\rm min}({\mathbf b})$ for which $\hat{\chi}^2$ is minimized as
\begin{equation}
        a_{\rm min}({\mathbf b}) = \frac{\sum_{i}\frac{(y - \sum_{
        j}b_{j}x_{ij})}{\sigma_{y_{i}}^2 + \sum_{j}(b_{
        j}\sigma_{x_{ij}})^2}}{\sum_{i}\left(\sigma_{y_{i}}^2 +
        \sum_{j}(b_{j}\sigma_{x_{ij}})^2\right)^{-1}}
\end{equation}

Finding the best fit parameters is then a simple minimization problem which
can be treated with standard numerical minimization routines.  Because our
analysis is restricted to a bivariate problem in this paper, it is possible
to use a graphical solution in the form of 2-D $\hat{\chi}^2$ contour plots.

Unfortunately, such a technique cannot handle censored
data. However, it is easy to verify that the incidence of the censored data
points in our sample on the estimate of the linear regression coefficient
is small compared to the intrinsic scatter in the data. In order to do so,
we have performed standard linear regression fit using specific techniques
that are able to handle censored data, as implemented in the ASURV package
\cite{ifn86}, on the whole data sample both with and without the upper
limits. The differences in the resulting multivariate linear correlation
coefficients in the two cases are much smaller than the estimated errors.

In the limit of $\sigma_{y_{i}} \gg \sigma_{x_{ij}}$, the merit function in
eq.~(\ref{eq:merit}) recovers the traditional $\chi^2$ estimator.  This
implies that the usefulness of eq.~(\ref{eq:merit}) is limited to cases
where the estimated errors are an accurate representation of the true
intrinsic uncertainty, including intrinsic scatter.  Artificially large
errors in one variable will introduce an asymmetry similar to that
intrinsic to the traditional regression formalism. At face value, the
errors in our sample are dominated by the uncertainties in the mass
measurements.  However, because we are using a sample of predominantly low
redshift low luminosity AGN, error in the distance measurements can be
large due to peculiar velocities, which leads to large errors in the
luminosity measurements as well.  For Galactic sources, distance estimates
typically carry even larger uncertainties.  Worse yet, the resulting errors
in $L_{\rm X}$ and $L_{\rm R}$ will be correlated.  As a first step, we
therefore decided to follow the Nuker approach and assume isotropic
uncertainties $\sigma_{M_{\rm i}}=\sigma_{L_{\rm X}, i}=\sigma_{L_{\rm R},
i}$ in all three variables and re-normalize these errors to produce a
minimum reduced $\hat{\chi}^2_{\rm red}=\hat{\chi}^2/n_{\rm dof}$ of unity.

\section{Results}
\label{sec:results}
We fit the data with the function
\begin{equation}
\label{eq:lr}
\log L_{\rm R}=\xi_{\rm RX}\log L_{\rm
  X} + \xi_{{\rm R}M}\log M + b_{\rm R},
\end{equation} 
or, alternatively with its reciprocal
\begin{equation}
\label{eq:lx}
\log L_{\rm X}=\xi_{\rm XR}\log L_{\rm
  R} + \xi_{{\rm X}M}\log M + b_{\rm X},
\end{equation}
where we have used the merit function to estimate the best fit linear
regression coefficients. Because the regression technique we adopt is
symmetric\footnote{We verified that the standard regression technique
  is instead not symmetric: for the same fitting function
  (\ref{eq:lr}), the best fit parameters are $\xi_{\rm RX}=0.57$ and
  $\xi_{\rm RM}=0.77$, while by fitting eq.~(\ref{eq:lx}) we obtain
  $\xi_{\rm XR}=1.05$ and $\xi_{\rm XM}=-0.49$, instead of the
  expected $\xi_{\rm XR}=1/\xi_{\rm RX}=1.75$ and $\xi_{\rm
    XM}=-\xi_{\rm RM}/\xi_{\rm RX}=-1.35$.}, eqs.~(\ref{eq:lr}) and
(\ref{eq:lx}) contain the same amount of information.

\begin{figure*}
\begin{center}
\psfig{figure=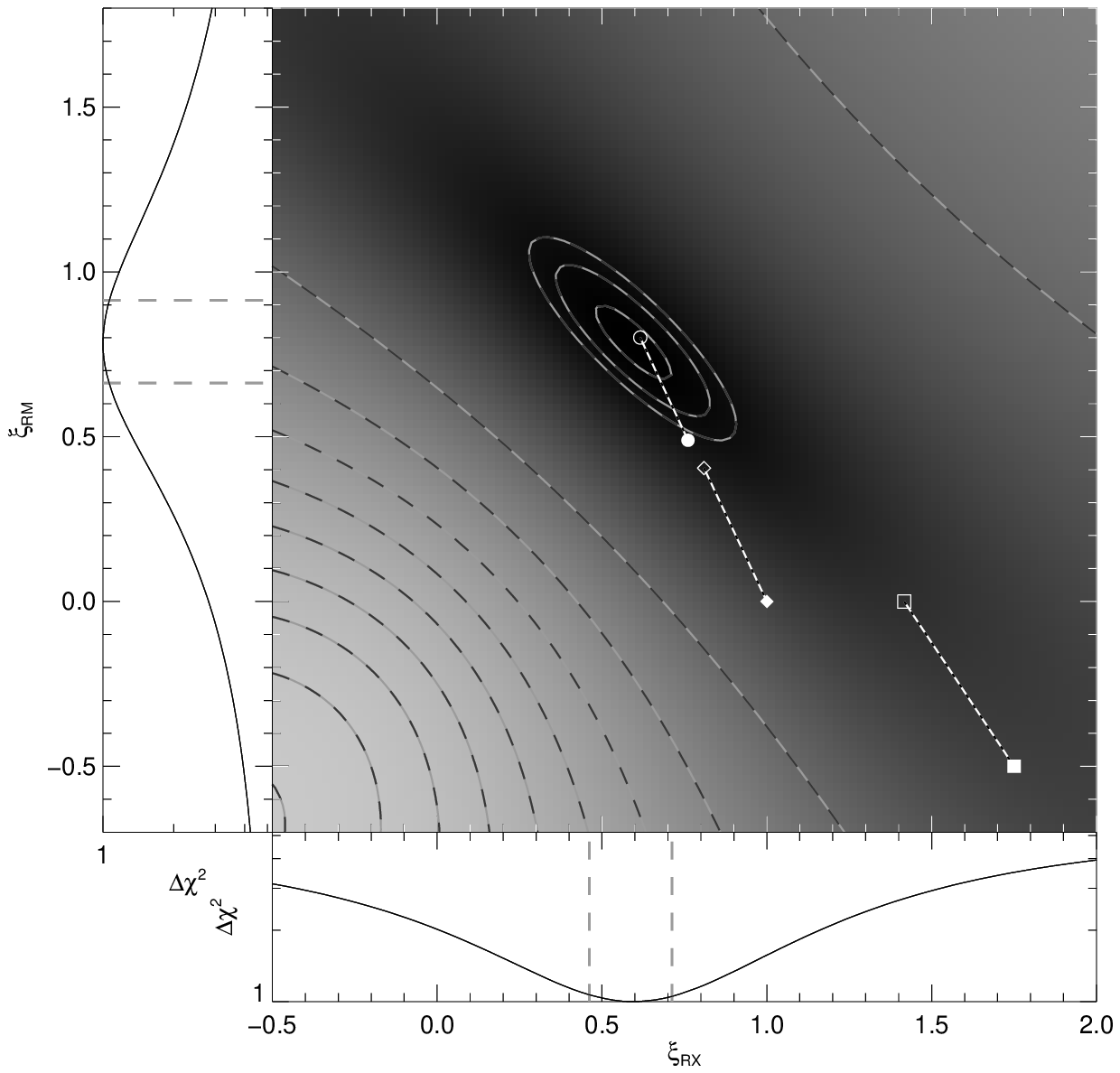,angle=0,width=0.77\textwidth}
\hbox{\psfig{figure=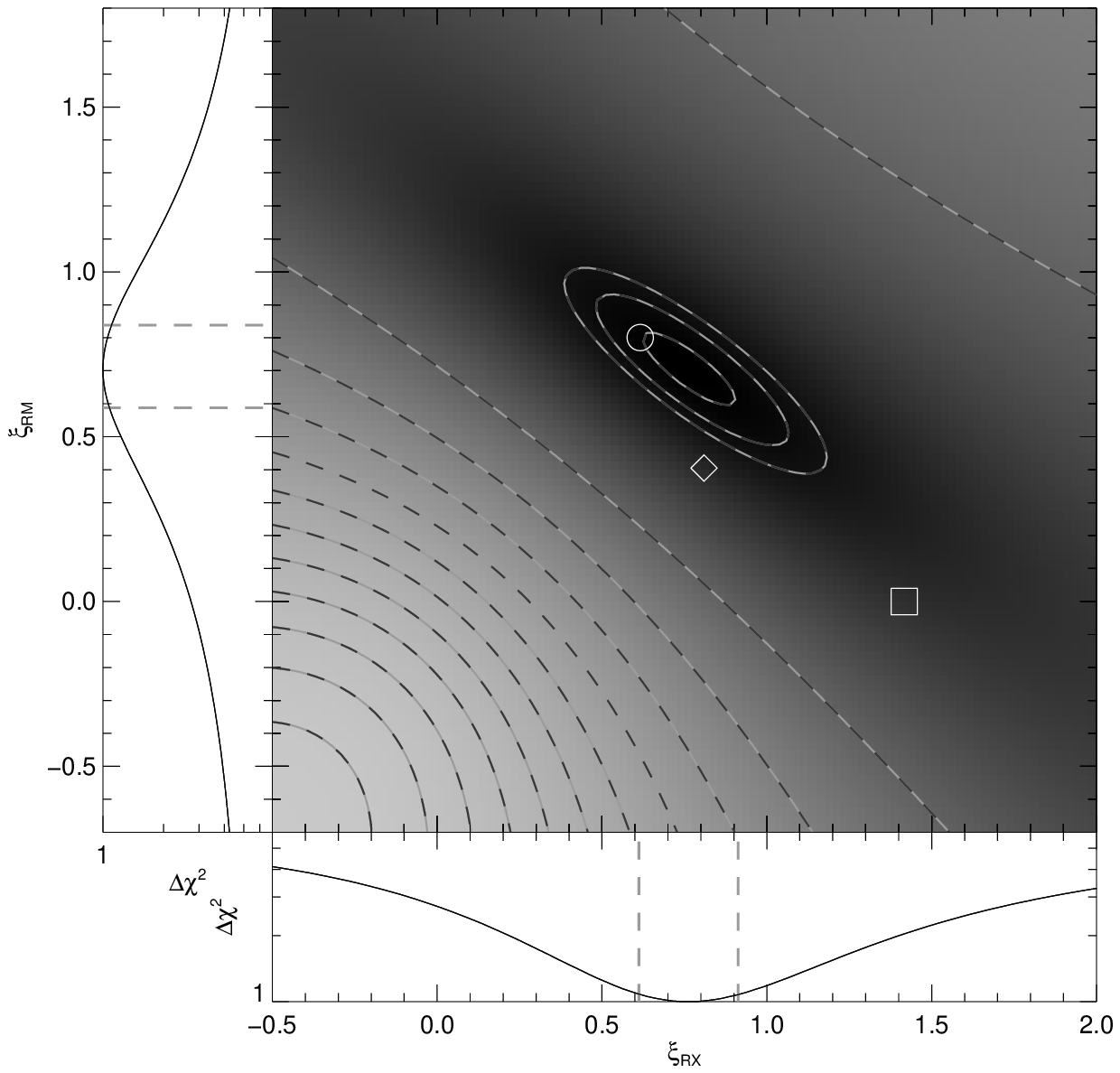,angle=0,width=0.4\textwidth}\hspace{1.3cm}
\psfig{figure=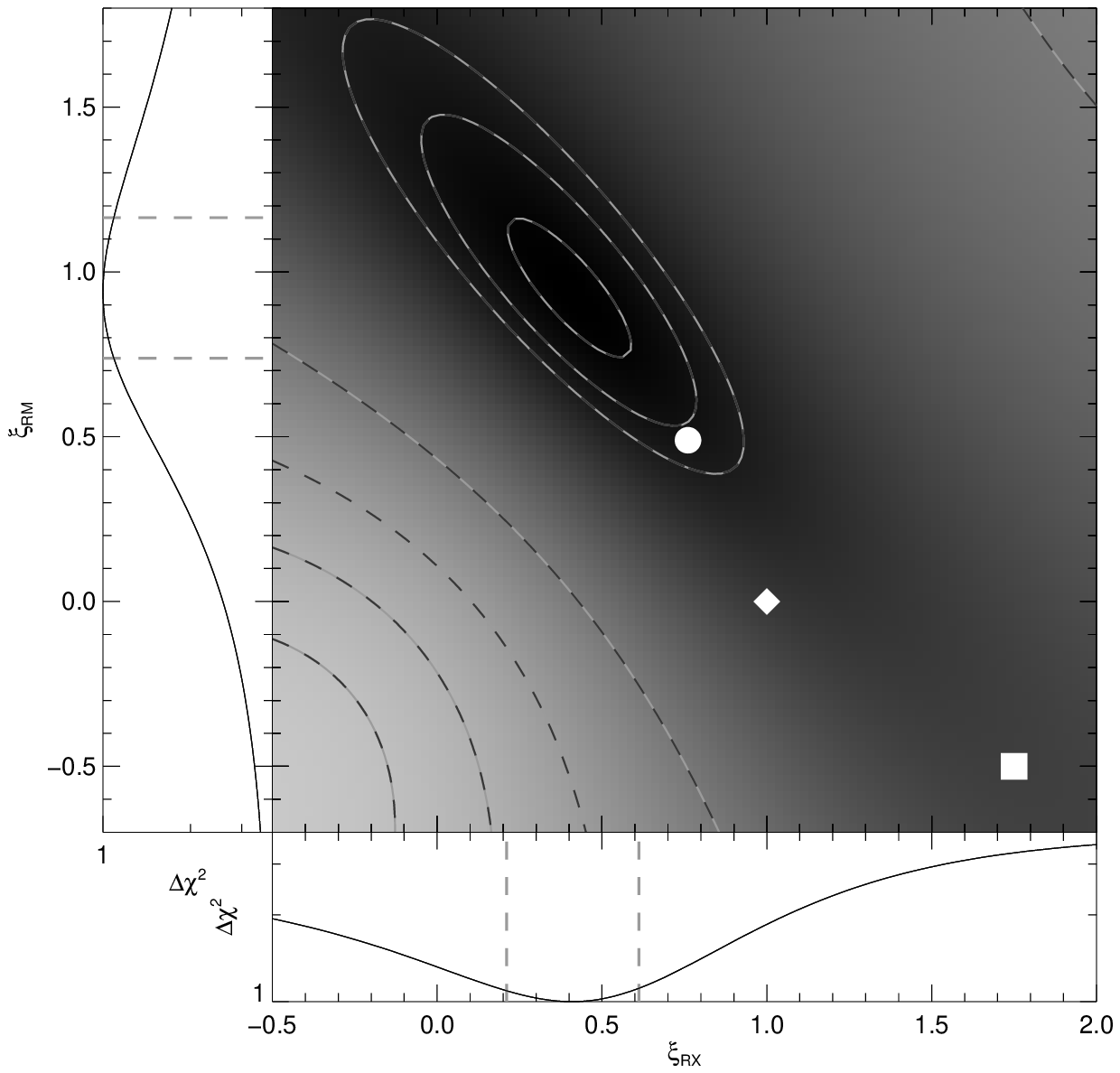,angle=0,width=0.4\textwidth}}
\end{center}
\caption{Shaded areas show the $\chi^2$ density distribution, and 
  dashed lines the $\chi^2$ contours for the observed correlation
  coefficients $\xi_{\rm RM}$ and $\xi_{\rm RX}$. The inner 3 contours
  show the formal 1,2 and 3 sigma confidence levels, the remaining
  contours further out show levels of $\Delta\chi_{\rm red}^2 = 10$.
  Shown are the results for: the entire data set (upper panel), the
  flat spectrum sources (lower left panel), the steep spectrum sources
  (lower right panel). Overplotted on each panel are the theoretically
  predicted values of the correlation coefficients where circles,
  diamonds and squares represent the ADAF, jet and standard disc
  models respectively. Empty symbols show the values for $\alpha_{\rm
    R} = 0$ and filled ones for $\alpha_{\rm R} =0.5$.  The lines connecting
  the points represent the tracks of $\xi_{\rm RX}$ and $\xi_{\rm RM}$
  traced out by variation of $\alpha_{\rm R}$.}
\label{fig:nukerstats1}
\end{figure*}

Our final results of the multivariate regression analysis are shown in
the form of $\chi^2$ contour plots (using eq.~(\ref{eq:merit}) as a
$\chi^2$ estimator) in the ($\xi_{\rm RM},\xi_{\rm RX}$) parameter
space in Fig.~\ref{fig:nukerstats1}. The top panel shows the results
of the multivariate fit performed on the entire data set. We obtain
$\xi_{\rm RX}=0.60 \pm 0.11$, $\xi_{\rm RM}=0.78^{+0.11}_{-0.09}$, and
$b_{\rm R}=7.33^{+4.05}_{-4.07}$ (all the errors are one sigma
confidence), with a dispersion $\sigma_{\rm R}=0.88$, or, cast in the form
of eq.~(\ref{eq:lr}):
\begin{equation}
\log L_{\rm R}=(0.60^{+0.11}_{-0.11}) \log L_{\rm X}
+(0.78^{+0.11}_{-0.09}) \log M + 7.33^{+4.05}_{-4.07}
\label{eq:fp}
\end{equation}

The value we obtain for the $\xi_{\rm RX}$ correlation coefficient is
consistent, within the errors, with that found in GBH 
($\xi_{\rm RX} \approx 0.7$) by Gallo et al. (2003). 
This also means that individual GBH sources for which the
correlation between radio and X-ray luminosities is well established
(GX 339-4 and V404 Cyg) do indeed follow the same global trend defined
by black holes of all masses included in our sample. 

\begin{figure*}
\psfig{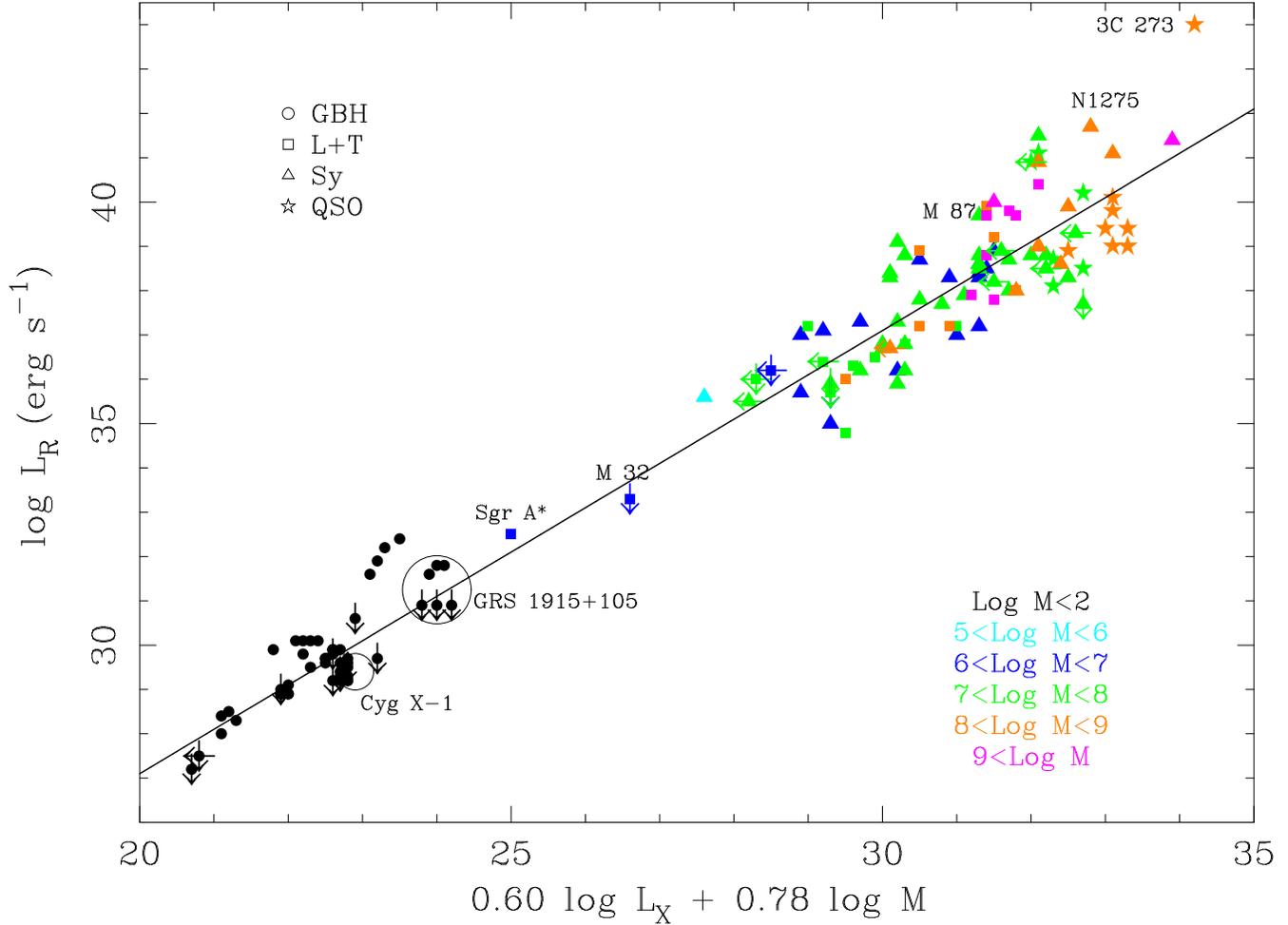}
\caption{The edge-on view of the ``fundamental plane of black hole
  activity''. The solid line shows the best fitting function (\ref{eq:fp}).}
\label{fig:plane}
\end{figure*}

Our data set provides fairly tight constraints on the values of the
correlation coefficients $\xi_{ij}$ ($i,j={\rm R,X,M}$). This in turn
implies that in the 3-dimensional space $(\log L_{\rm R},\log L_{\rm
  X},\log M)$ the sources are distributed preferentially on a plane,
that we call hereafter the ``fundamental plane'' describing black hole
activity. Figure~\ref{fig:plane} shows an edge-on view of the plane.

The other two panels of Fig.~\ref{fig:nukerstats1} show the
constraints on the two correlation coefficients obtained with the
regression analysis on different subsamples assembled according to the
spectral properties in the radio band.  For flat spectrum radio
sources we obtain $\xi_{\rm RX,f}=0.76 \pm 0.13$, $\xi_{\rm RM,f}=0.71
\pm 0.10$, and $b_{\rm R,f}=1.31^{+4.85}_{-5.07}$, with a dispersion
$\sigma_{\rm R,f}= 0.81$, smaller than that of the whole data set.
The difference in the coefficient for the $L_{\rm R}-L_{\rm X}$
correlation with respect to the whole sample result may be due to a
larger incidence of GBH in the flat spectrum subsample, which seem to
have a slightly larger measured value of $\xi_{\rm RX}$ \cite{gfp02};
the constraints on the other coefficient, $\xi_{\rm RM}$ are almost as
good as for the entire data set, and in very good agreement with it.
 
On the other hand, in our steep spectrum sources subsample, only one
GBH is included (LS 5039), and this results in a much larger
uncertainty, in particular on the radio luminosity-mass correlation
coefficient: ($\xi_{\rm RX,s}=0.41^{+0.16}_{-0.18}$, $\xi_{\rm
  RM,s}=0.94^{+0.21}_{-0.19}$, $b_{\rm R,s}=14.08^{+5.92}_{-6.31}$
with dispersion $\sigma_{\rm R,s}=0.87$).

\section{Physical interpretation of the fundamental plane relation}
\label{sec:physics}

 As stated in the introduction, a correlation between X-ray and radio
emission is expected if there is a fundamental connection between accretion
flows and jet activity.  Thus, at a qualitative level, the existence of the
fundamental plane found in \S\ref{sec:results} is not surprising.  However,
on a quantitative level, the presence of such a plane and the measurement
of the correlation coefficients associated with it provide powerful probes
of accretion physics and of the inner jet.

In the following section we will therefore lay out an avenue of how to use
the fundamental plane relation to constrain accretion and jet physics.  In
order to do this, we shall first turn to a discussion of the theory of the
radio and of the X-ray emission by the jet--disc system.  We will first
make use of the scale invariant assumption about the disc--jet coupling
\cite{hs03} to derive the expected scalings of the radio luminosity with
black hole mass and accretion rates. Then, we will show how these
theoretically predicted correlation coefficients translate into observable
ones between black hole mass and X-ray luminosity ($\xi_{\rm RM},\xi_{\rm
RX}$) for different models of the X-ray emission.  By comparing these
theoretical predictions to the observed values from eq.~(\ref{eq:fp}) we
will then discuss the constraints we can put on the models themselves.

\subsection{Synchrotron emission from scale invariant jets}
\label{sec:sync}
It was recently shown by Heinz \& Sunyaev (2003) that the dependence of
radio luminosity $L_{\rm R}$ on black hole mass $M$ and dimensionless
accretion rate $\dot{m}$ can be cast into a model independent form {\em if
the underlying jet physics is scale invariant}.  In this case, all the
model dependent uncertainties can be absorbed into the observable spectral
index $\alpha$.  The relationships between $L_{\rm R}$ and $M$ and between
$L_{\rm R}$ and $\dot{m}$ depend only on the boundary conditions at the
base of the jet, set by the conditions in the accretion flow feeding the
jet.  We will briefly review their argument (the interested reader is
encouraged to consult Heinz \& Sunyaev, 2003, for more details) before
comparing their predictions to the correlations derived in the previous
section.

The nature and conditions in the inner disc are most likely governed by
only a few parameters ($M$, $\dot{m}$, and possibly the black hole spin
$a$).  The fundamental scale imposed on the problem is the gravitational
radius of the black hole, $R_{\rm g}=GM_{\rm BH}/c^2$.  Jet formation
occurs in the innermost regions of accretion discs, thus it is natural that
jet formation, too, is primarily governed by these three
parameters only\footnote{The influence of $a$ and the associated second scale of
the problem, the light cylinder radius, on jet formation is unclear at this
point.  We will henceforth keep $a$ fixed and assume that variations in $a$
will only introduce a scatter in any relation derived below which is
independent of $M$ and $\dot{m}$.}.

Then, we make the {\it ansatz} that jet structure and dynamics are (at least
approximately) invariant under changes of $R_{\rm g}\propto M$ and
$\dot{m}$.  In other words, taking a jet produced by a black hole of mass
$M_1$, scaling its dimensions by a factor of $M_2/M_1$ 
we should obtain a jet around a black
hole of mass $M_2$.

This proposed invariance can be cast into a simple mathematical form. Any
quantity $f$ needed to calculate the synchrotron emission from jets can be
decomposed into a structure function $\psi_{f}({\mathbf R}/R_{\rm g},a)$,
which describes the spatial variation of $f$ along the jet, and which
depends on $M$ only through ${\mathbf R}/R_{\rm g}$, and a normalization
$\phi_{f}(M,\dot{m},a)$, which is set by the boundary conditions at the
base of the jet (thus, by accretion disc physics):
\begin{equation}
\label{eq:sc_inv}
        f({\mathbf R},M,\dot{m},a) = \phi_f(M,\dot{m},a) \cdot
        \psi_f({\mathbf R}/R_{\rm g},a)
\end{equation}

The quantities necessary for calculating the jet synchrotron emission are
the magnetic field strength $B$, the jet diameter $D_{\rm jet}$, and the
normalization $C$ of the electron power-law distribution $dn_{\rm
e}/d\gamma = C\gamma^{-p}$ (typically, the spectral index of the particle
distribution is $p\sim 2-3$).  For example, according to the prescription
(\ref{eq:sc_inv}), the magnetic field should follow $B =
\phi_{B}(M,\dot{m},a)\cdot \psi_{B}({\mathbf R}/R_{\rm g},a)$, where
$\phi_B=B_0=B(R_0)$ is the value of the field at the base of the jet $R_0$.
As a geometric quantity, the jet diameter $D$ should be directly
proportional to the characteristic scale $R_{\rm g}$ such that
$\phi_D=D_0=D(R_0)\propto R_{\rm G}\propto M$.  
Scale invariance also implies that dynamical time scales are proportional
to $R_{\rm g}/c\propto M$, and thus that characteristic velocities are
scale invariant.  

Using the standard formulae for synchrotron emission \cite{rl79}, Heinz \&
Sunyaev (2003) showed that the synchrotron luminosity $L_{\nu}$ at a given
frequency $\nu$ emitted by the jet must then depend non-linearly on $M$ and
$\dot{m}$, following
\begin{eqnarray}
\lefteqn{\frac{\partial \ln{(L_{\nu})}}{\partial \ln{(M)}} =
        \frac{2p+13+2\alpha}{p+4} +
        \frac{\partial\ln{(\phi_{B})}}{\partial\ln{(M)}}
        \left(\frac{2p+3+\alpha p + 2\alpha}{p+4}\right)} \nonumber \\ & &
        \ \ \ \ \ \ \ \ \ \ \ \ \ + \ \
        \frac{\partial\ln{(\phi_{C})}}{\partial\ln{(M)}}
        \left(\frac{5+2\alpha}{p+4}\right) \equiv \xi_{M} \label{eq:mnu}
\end{eqnarray}
and
\begin{eqnarray}
\lefteqn{\frac{\partial \ln{(L_{\nu})}}{\partial
        \ln{(\dot{m})}} = 
        \frac{\partial\ln{(\phi_{B})}}{\partial\ln{(\dot{m})}}
        \left(\frac{2p+3+\alpha(p + 2)}{p+4}\right)} \nonumber \\ & &
        \ \ \ \ \ \ \ \ \ \ \ \ \ + \ \
        \frac{\partial\ln{(\phi_{C})}}{\partial\ln{(\dot{m})}}
        \left(\frac{5+2\alpha}{p+4}\right) \equiv \xi_{\dot{m}}
        \label{eq:mdotnu}
\end{eqnarray}
where $\alpha$ is the spectral index at frequency $\nu$.

Note that 
the model dependent structure functions $\psi_f({\mathbf R}/R_{\rm g},a)$
scale out from these expressions.  Only the spectral indices
($\alpha$ and $p$) and the boundary conditions $\phi_B$ and $\phi_C$ for
the magnetic field $B$ and the electron power-law distribution
normalization $C$, respectively, remain.  $\alpha$ and $p$ are observables:
the electron spectral index $p$ can be deduced from the optically thin
synchrotron spectral index at high frequencies. The functions $\phi_B$
and $\phi_C$, on the other hand, need to be provided by accretion disc
theory. It is reasonable to
assume that the relativistic particle pressure at the injection radius is a
fixed fraction (i.e., independent of $M$ and $\dot{m}$) of the total
pressure at injection, $\phi_{C}\propto \phi_{B}^2$.  This leaves
$\phi_B$ as the only model dependent parameter of the theory. Given a
prescription for $\phi_B$, we can predict how the synchrotron luminosity of
jets should scale with $M$ and $\dot{m}$.

Since $\xi_{M}$ and $\xi_{\dot{m}}$ are constants, we have in general (with
$L_{\nu}=L_{\rm 5\,GHz}=L_{\rm R}$)
\begin{equation} 
        \log L_{\rm R}=\xi_M \log M + \xi_{\dot m} \log \dot m + K_1,
        \label{eq:logr}
\end{equation}
where $K_1$ is a normalization constant\footnote{We note here that for flat
spectrum jets with $\alpha_{\rm R}\sim 0$, the canonical value of $p\sim
2$, and $\phi_{B}^2 \propto M^{-1} \dot{m}$, the dependence of $L_{\rm R}$
on $M$ and $\dot{m}$ follows $L_{\rm R}\propto
\left(M\dot{m}\right)^{17/12} =\dot{M}^{17/12}$, as had been found by
Falcke \& Biermann (1996) for the specific case of the ``canonical
conical'' \cite{bk79} jet model.}. Variations in other parameters, such as
the viewing angle or the black hole spin, will only introduce a scatter in
this relationship that is {\em independent} of $M$ and $\dot{m}$.

Because the expressions for $\xi_{M}$ and $\xi_{\dot{m}}$ do not depend on
the shape functions $\psi_f$, they are independent of the model details.
Any scale-invariant jet model that reproduces the observed radio spectral
index $\alpha_{\rm R}$ {\em must} satisfy eqs.~(\ref{eq:mnu}) and
(\ref{eq:mdotnu}).  This implies that
(a) measurements of $\xi_{M}$ and $\xi_{\dot{m}}$ {\em cannot} be used
to constrain the functions $\psi_f$, i.e., they cannot be used to
distinguish between different jet models but that (b) measurements of
$\xi_{M}$ and $\xi_{\dot{m}}$ {\em can} be used to place constraints on the
boundary conditions at the base of the jet $\phi_f$, or in other words
on the accretion disc model.

While the black hole mass $M$ is observable (see above), the accretion rate
$\dot{m}$ can only be inferred through radiation in other bands, where the
emission is dominated by the accretion disc.  Thus, in addition to a
prescription of the boundary conditions $\phi_f$ through accretion disc
theory, we must also provide a relation between the accretion disc
luminosity (or $L_{\rm X}$) and $\dot{m}$ in order to model the observed
$L_{\rm R}$-$\dot{m}$ correlation.  We must therefore discuss the different
possible sources of the X-ray emission. We will start with those models
that describe X-ray emission as produced by an accretion flow of some kind.

\subsection{Accretion flow origin of X-ray radiation}
\label{sec:acc}

For the accretion powered X-ray luminosity we can write, in analogy to
eq.~(\ref{eq:logr}):
\begin{equation}
\log L_{\rm X}= \log M + q \log \dot m  + K_2
\label{eq:logx}
\end{equation}
where $K_2$ is a normalization constant.  The efficiency coefficient $q$
need not be constant, though the linear correlation analysis from
\S\ref{sec:corr} is limited to this case.

Using eqs.~(\ref{eq:mnu}-\ref{eq:logr}) together with eq.~(\ref{eq:logx}),
we obtain the following general expression for the {\it observable}
correlation coefficients as defined in eqs.~(\ref{eq:lr}) and
(\ref{eq:lx}):
\begin{eqnarray}
\label{eq:observablexi}
\xi_{{\rm R}M} &=& \frac{2p+13+2\alpha_{\rm R}}{p+4}+
\frac{\partial \ln \phi_B}{\partial \ln
  M}\left(\frac{2p+13+\alpha_{\rm R} p
    +6\alpha_{\rm R}}{p+4}\right) \nonumber \\ &-& \frac{\partial \ln \phi_B}{\partial \ln \dot m}
\left(\frac{2p+13+\alpha_{\rm R} p
    +6\alpha_{\rm R}}{q(p+4)}\right)
\nonumber\\
\xi_{\rm RX} &=& \frac{\partial \ln \phi_B}{\partial \ln \dot m}
\left(\frac{2p+13+\alpha_{\rm R} p
    +6\alpha_{\rm R}}
{q(p+4)}\right).
\end{eqnarray}
Different accretion models imply different values of $q$, and different
scalings of the magnetic energy density at the base of the jet $\phi_B$
with $M$ and $\dot m$, resulting in different values of $\partial \ln
\phi_B/\partial \ln M$ and $\partial \ln \phi_B/\partial \ln \dot m$. 
In the next two sections we calculate the expected scalings for
radiatively efficient and inefficient flows 
(the results of the calculations are
summarized in Table~\ref{tab:disks}). 
For the case of radiatively efficient accretion, we will consider the
X-ray emission to be produced by a standard, geometrically thin and
optically thick disc coupled with a hot corona.
In the case of radiatively inefficient accretion flows, 
for the sake of clarity
we will describe a detailed spectral modeling 
in the framework of a pure advective disk (ADAF). Currently this 
 is the simplest existing dynamical model 
of a radiatively inefficient flow  from which we can derive detailed
spectral energy distributions. However, one should keep
in mind that radiatively inefficient 
flows may be significantly modified by convection
or outflows (see below). We do not, however, 
expect the arguments laid out below to be significantly modified in such cases.

As a result of our calculations, we will show how the physical
properties of radiatively inefficient and efficient discs do indeed
imply very different scalings for the observed X-ray luminosity with
the accretion rate $\dot m$.

\subsubsection{Radiatively efficient flows}
\label{sec:corona}
In the standard Shakura-Sunyaev disc model \cite{ss73} accretion occurs via
an optically-thick and geometrically thin disc. 
The effective optical depth
in the disc is very high and photons are close to thermal equilibrium with
electrons. The emission results in multi-color ($T \propto \dot m
^{1/4} M^{-1/4} r
^{-3/4}$) blackbody spectrum. 
This component is thought to explain the
``blue bump'' in AGN and the soft X-ray emission in GBH. However, the
standard model does not predict the power-law X-ray emission observed in
all sub-Eddington accreting black holes.
In fact, the emission in the 2-10 keV band we are considering here 
is completely dominated by the power-law
spectral component for SMBH and for GBH in the low/hard state.

It is generally accepted that this hard X-ray power-law 
emission is produced by inverse
Compton scattering of the soft, blackbody disc photons on a population of
hot electrons that surround the cooler disc in the innermost region of the
accretion flow (the so-called {\it corona})
In these models a
fraction $f_{\rm c}$
of the gravitational power is dissipated in the corona (and eventually
emerges as X-ray radiation).  The X-ray luminosity can then be written as
$L_{\rm X} \sim f_{\rm c} \dot m L_{\rm Edd} \propto f_{\rm c} \dot m
M$, 
and the coronal magnetic energy density follows 
$B^2 \propto \dot m f_{\rm c}/M$ (e.g.; Haardt \&
Maraschi 1991; Di Matteo, Celotti \& Fabian 1999; Merloni \& Fabian 2002).
$L_{\rm X}$ should therefore scale linearly with $\dot{m}$, as long as the
fraction of power dissipated into the corona $f_{\rm c}$ 
does not change with the
accretion rate.
Note, however, that we expect some scatter in any such a 
relation due to intrinsic changes in the coronal plasma
optical depth and temperature that may have some dependence on the
accretion rate.

Merloni \& Fabian (2002) and Merloni (2003) have shown how it is possible
to derive coupled equations for the accretion disc--corona systems under
very simple assumptions about the nature of the turbulent magnetic
viscosity inside the disc. The main property of their self-consistent
coupled treatment is that $f_{\rm c}$ is indeed constant when gas pressure
dominates in the disc, so for accretion rates $\dot m \la 0.016
(\alpha_{\rm v} M)^{-1/8} (1-0.84 \alpha_{\rm v}^2)^{-9/8}$ (with
$\alpha_{\rm v}$ viscosity parameter) we should expect $q=1$. On the
other hand, $f_{\rm c}\propto \dot
m^{-1/2}$ in the radiation pressure dominated parts of the disc, and
$q=1/2$ (see Table~\ref{tab:disks}). 
It is worth noting that $q$ and
$\partial \phi_{B}/\partial \dot{m}$ enter eq.~(\ref{eq:observablexi}) in
such a way that the correlation coefficients $\xi_{\rm RM}$ and $\xi_{\rm
RX}$ are identical for the gas and radiation pressure dominated disc cases.


\subsubsection{Radiatively inefficient flows}
\label{sec:adaf}

The standard solution we have discussed in the previous section is not a
unique solution for the accretion flow equations at low accretion rates.
In recent years, much work has been devoted to the detailed
study of low radiative efficiency accretion and in particular to
the so called Advection Dominated Accretion Flows (ADAF) solutions
\cite{ny94,ny95,abr95,nmq98}.  
The latter usually refers to the optically--thin ADAF branch 
which is established only for accretion rates lower than
a critical value $\dot m < \dot m_{\rm crit} \sim \alpha_{\rm v}^2$ 
\cite{ree82}.
However, both from the theoretical point of view
\cite{nia00} and from numerical simulations \cite{ia00,hb02}, it has
been shown that that radiatively inefficient flows are prone to strong
convective instabilities and/or powerful outflows which alter the
nature of the solutions significantly.  Despite the extensive
theoretical efforts, however, the relative importance of convection
and outflow for adiabatic flows is still a matter of significant
debate \cite{bh02,nq02} and no strong observational discriminant has
yet been found.


From our point of view, it is important to understand how the
value of the magnetic field in such flow scales with $M$ and $\dot m$, and
 to have a detailed model for their radiative output, in order to
predict the scaling of the X-ray emission with these parameters.  From
general arguments, we expect any mechanically cooled flow to obey
$\phi_{B}^2 \propto M^{-1} \dot{m}$ \cite{hs03}. On the other hand, the
exact value of $q$ can only be determined once a detailed assessment of the
radiative processes that give rise to observed luminosity is made. This is
what we discuss in the following. 

A hot accretion flow around a stellar mass (supermassive) black hole
radiates mostly in the optical (radio) to X--ray bands. In the optical
(radio) band, the emission results from synchrotron radiation.  At higher
energies, and up to the X-ray band, the emission is produced by
bremsstrahlung processes for low accretion rates and inverse Compton
scattering of the soft synchrotron photons or blackbody photons from the
disc in the outer regions when the accretion rate approaches the critical
value \cite{nbm97}.  The predicted spectrum from an ADAF depends (weakly)
on the ratio of the gas to magnetic pressure $\beta$, the viscosity
parameter $\alpha_{\rm v}$, and the fraction of the turbulent energy in the
plasma which heats the electrons, $\delta$. Here, we fix $\alpha_{\rm v} =
0.1$, $\beta = 10$, and take $\delta = 0.3$.

\begin{figure}
\begin{tabular}{c}
\psfig{figure=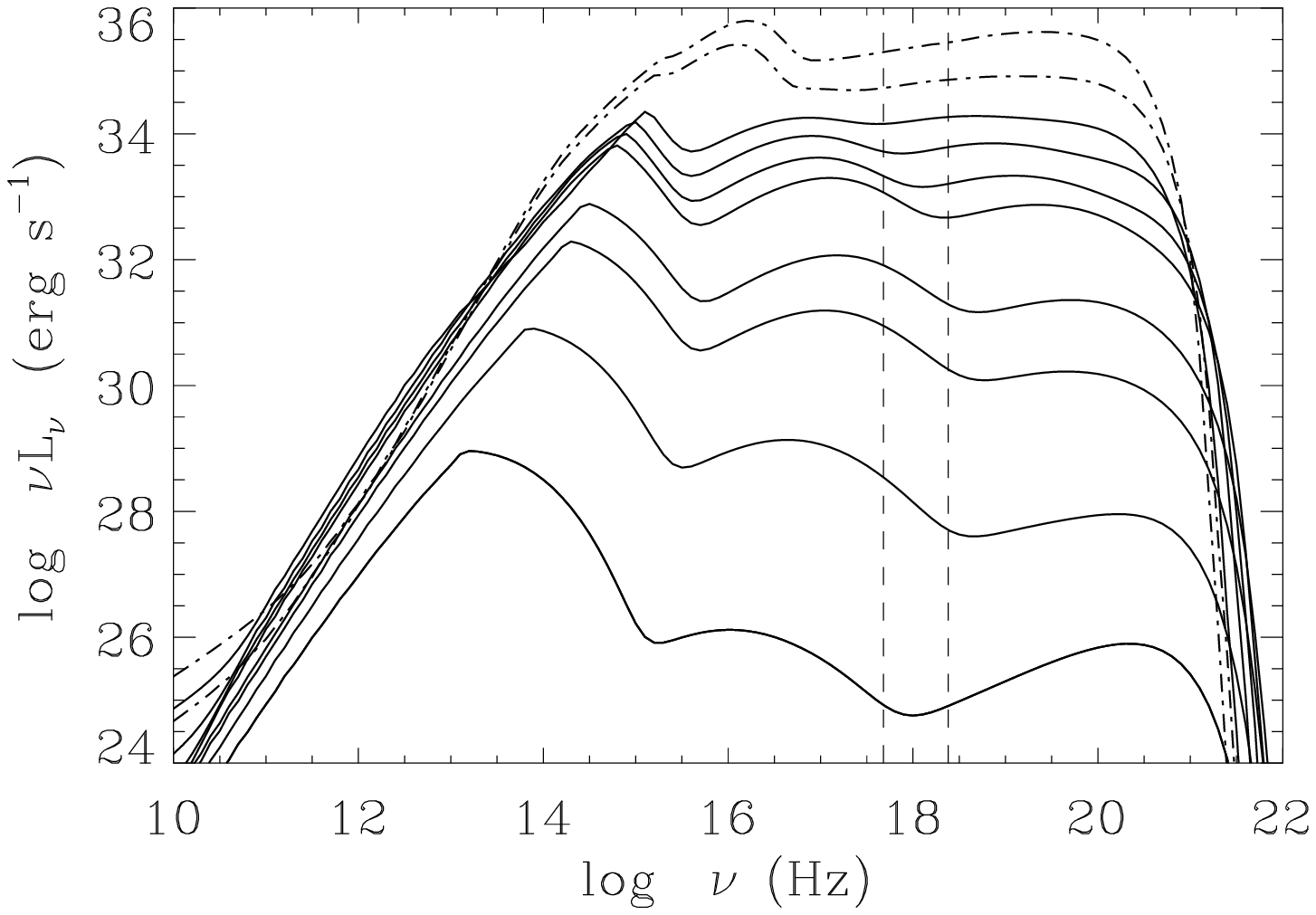,angle=0,width=0.5\textwidth}\\
\psfig{figure=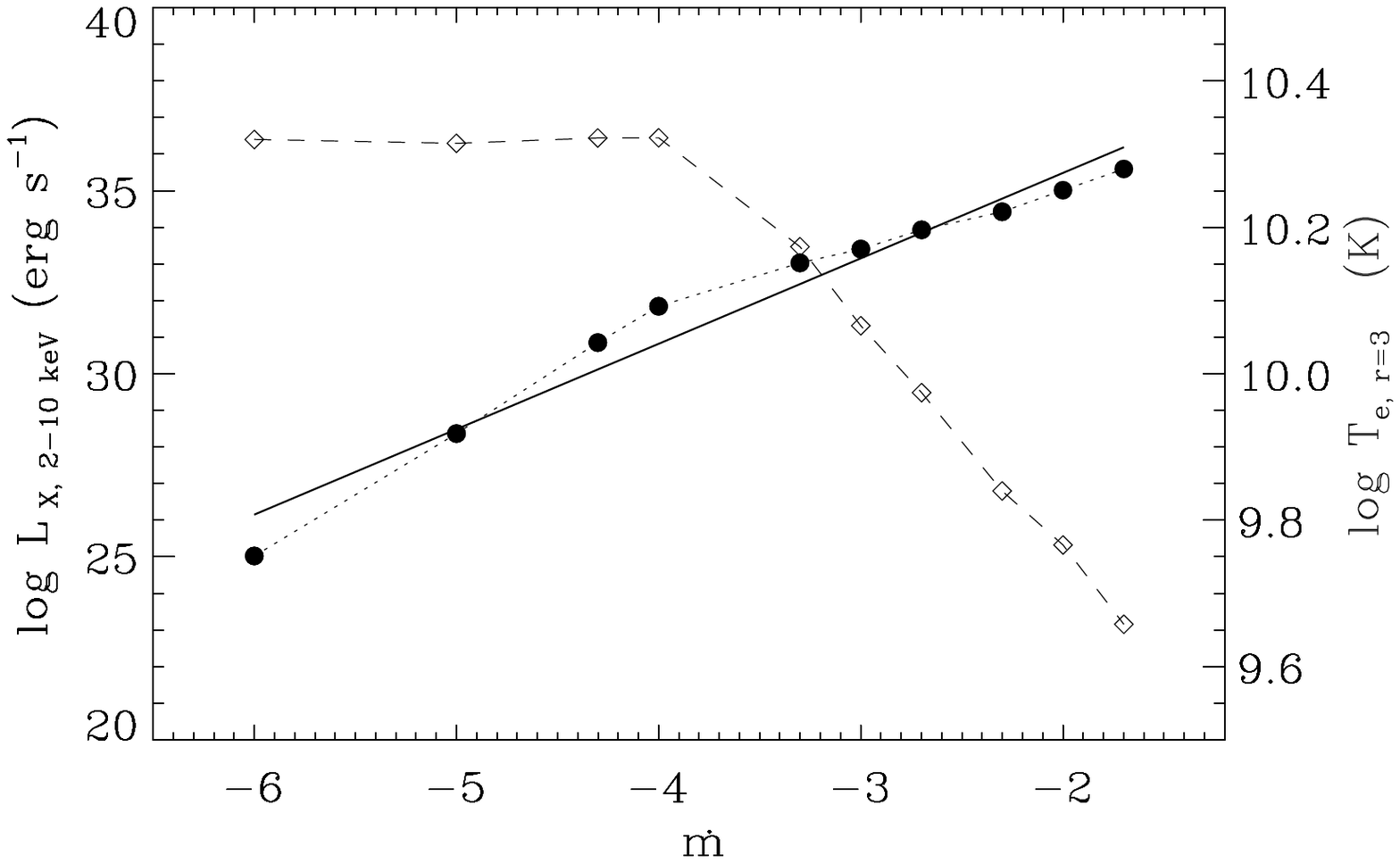,angle=0,width=0.5\textwidth}\\
\end{tabular}
\caption{Predictions from ADAF models for $\dot{m}= 10^{-6},10^{-5},5 \time
10^{-5},10^{-4},5\times 10^{-4}, 10^{-3}, 2\times 10^{-3},5\times 10^{-3},
10^{-2}$. The top panel shows the broad band spectral energy distributions
for a fixed $M=10$, with the vertical dashed lines enclosing the $2-10$ keV
spectral energy band. For the top two models (dot-dashed lines) we also
plot the multicolor blackbody from an outer thin disc with an inner edge at
$r=40$. In the bottom panel: the solid dots joined by the dotted line show
the integrated $2-10$ keV X-ray luminosity and the open symbols joined by
dashed lines and the flow electron temperature at $r=3$ versus the
accretion rate $\dot {m}$. The solid line is a linear fit to the $2-10$ keV
luminosity over the whole of the $\dot {m}$ with slope 2.3. }
\label{fig:adaf_lxm}
\end{figure}

The top panel of Figure~\ref{fig:adaf_lxm} shows the spectra of ADAFs (plus
outer thin disc for $\dot m \sim \dot m_{crit}$) for ten accretion rate
values starting from $\dot m = 10^{-6}$ up to $\dot{m} \sim \dot {m}_{crit}
\sim 10^{-2}$. The bottom panel shows the relationship between the $2-10$
keV luminosity, $L_{\rm X}$, and the accretion rate, $\dot{m}$ (for
$M=10$).  For the chosen values of the parameters $\alpha_{\rm v}$, $\beta$
and $\delta$, the dependence of the 2-10 keV luminosity on $\dot{m}$ is
roughly given by (see also Beckert \& Duschl 2002, for a similar calculation):
\begin{equation}
        L_{X, \rm 2-10 keV}  \propto \left\{
        \begin{array}{ll}
        \dot{m}^{3.4}, & \hbox{$ \dot{m} \la 10^{-4}$} \\
        \dot{m}^{1.6},   & \hbox{ $10^{-4} \la \dot{m} \la 5 \times
        10^{-3}$}   \\
        \dot{m}^{2},   & \hbox{ $5\times 10^{-3} \la \dot{m} \la 2 \times
        10^{-2}$}   \\
        \end{array}\right.\;
\label{eq:lxm}
\end{equation}

The full band spectra shown in Figure~\ref{fig:adaf_lxm} with the two
vertical lines indicating the $2-10$ keV energy band illustrate the origin
for the various dependences shown in equation (\ref{eq:lxm}).  For $\dot{m}
\la 10^{-4}$ the integrated $2-10$ keV emission includes both the first
order Comptonized component, which drops off at around few keV, and
bremsstrahlung which dominates the emission at higher energies.  At higher
accretion rates the first (or higher) order Compton scattering of softer
photons always dominate the $2-10$ keV emission (see
Figure~\ref{fig:adaf_lxm}, top panel).  At these higher rates the optical
depth increases and cooling becomes overall more efficient.

This can also been seen in bottom panel of Figure~\ref{fig:adaf_lxm}, where
we also plot (dashed line) the electron temperature at $r=3$ as a function
of $\dot{m}$. The temperature is nearly constant for $\dot{m} \la 10^{-4}$; in
this regime the viscous heating is virtually fully balanced by advection.
At $\dot{m} \ga 10^{-4}$ cooling processes become more important and the
temperature decreases.  The overall change in the dependences in
eq.~(\ref{eq:lxm}) reflects this behavior.  

However, it is important to keep in mind that the exact dependences in
this narrow band will also depend on the microphysical parameters that
we choose for the model.  For instance, variations in the electron
heating parameter, $\delta$ will result in changes in the
Comptonization spectra. Therefore, in order to compare theoretical
predictions with the observed data, the intrinsic scatter of which
does not allow us to put tight constraints on the different emission
regimes, we fit the $L_{\rm X} - \dot m$ relation over the whole range
of $\dot m$ and obtain a single power-law, $L_{\rm X} \propto \dot
m^{2.3}$ (i.e. $q=2.3$, as shown by the solid line in the 
middle panel of Figure
\ref{fig:adaf_lxm}).  We also calculate the relation between $L_{\rm
  2-10\,keV}$, versus the black hole mass, $M$ for the models and, as
expected, the relation is very close to linear with $L_{\rm X} \propto
M^{0.97}$. Overall the scalings derived from our specific computation
of a specific ADAF model are consistent with the expected behaviour of
the more general class of radiatively inefficient, 
mechanically cooled accretion flows. In fact, we note here that if the flow is
modified by convection or outflows, we expect the X-ray spectrum to be
dominated by  bremsstrahlung emission \cite{qn99,dm99}, 
for which we expect $q=2$.

\subsection{Jet origin of X-ray radiation}
\label{sec:xrayjet}
The presence of both high energy electrons and radio/optical radiation
implies that the inner jet must emit inverse Compton X-rays, contributing
to the overall X-ray spectrum at some level.  However, under suitable
conditions even the synchrotron component will reach X-ray energies.
Especially in the presence of relativistic beaming, synchrotron emission
could actually be responsible for the bulk of the X-rays.  It is therefore
useful to discuss the radio-X-ray-mass correlation expected for the
synchrotron X-ray scenario \cite{mff01}.

While at low frequencies the effects of synchrotron self-absorption lead to
the flat spectrum typically observed in core dominated sources, at high
frequencies the jet becomes optically thin to synchrotron self-absorption,
following the well know optically thin power-law with X-ray spectral index
$\alpha_{\rm X}=(p-1)/2$.

Because X-rays and radio emission now originate from the same source, it is
straightforward to derive the relation between the two components.  The
correlation coefficients $\xi_{M}$ and $\xi_{\dot{m}}$ for the optically
thick radio emission are still given by eqs.~(\ref{eq:mnu}) and
(\ref{eq:mdotnu}), while for the optically thin X-ray emission we can
simply substitute $\alpha_{\rm X}=(p-1)/2$ in those expressions.  Finally,
we can eliminate $\log{\dot{m}}$ from eq.~(\ref{eq:logr}) in favor of
$\log{L_{\rm X}}$ and arrive at the desired expressions for the correlation
coefficients:
\begin{eqnarray}
\label{eq:tildexi}
\tilde \xi_{\rm RX}&=&\frac{2(2p+13+\alpha_{\rm R} p+6\alpha_{\rm
    R})}{(p+4)(p+5)} 
\nonumber \\ \tilde \xi_{\rm
RM}&=&\frac{2p+13+2\alpha_{\rm R}}{p+4}-\frac{6(2p+13+\alpha_{\rm R}
p+6\alpha_{\rm R})}{(p+4)(p+5)}
\end{eqnarray}
where $\alpha_{\rm R}$ is the radio spectral index and $p$ is assumed to be
the same for radio and X-ray emitting electrons\footnote{This implies that
radiative cooling must either be negligible for the region of the jet where
the X-rays are produced, leaving $p\sim 2$ over the entire spectrum, or
that cooling and continuous injection have produced a universal power-law
with a slope of order $p\sim 3$.  This limitation, while severe, cannot be
avoided because particle transport including the effects of radiative
cooling cannot be formulated in a simple scale invariant fashion.}, and we
have used a tilde to distinguish the synchrotron X-ray model.  


\begin{table}
\caption{Scaling indices of the X-ray luminosity and of the magnetic
  field as functions of the physical parameters $\dot m$ and $M$ for
  different accretion flow models.}
\label{tab:disks}
\begin{tabular}{lccc}
\hline
\hline
accretion model & $q$ &$\frac{\partial \ln \phi_B}{\partial \ln M}$ 
&$\frac{\partial \ln \phi_B}{\partial \ln \dot m}$ \\
\hline 
disc/corona, gas & 1 & -1/2  & 1/2\\
disc/corona, rad & 1/2 & -1/2  & 1/4\\
ADAF & ~2.3 & -1/2  & 1/2\\
brehms. emission only & 2 &  -1/2  & 1/2\\
\hline
\end{tabular}
\end{table}

\subsection{Comparison with observations}
\label{sec:montecarlo}
In the previous sections we have shown how different theoretical
models for the emission processes responsible for the observed radio
and X-ray emission from black holes can be directly translated into
predictions for the observable correlation coefficients $\xi_{\rm RX}$
and $\xi_{\rm RM}$ (see eq.~\ref{eq:lr}). The relatively tight
observational constraints on these indices that we have derived in
section~\ref{sec:corr} can therefore be used to directly test these
models. The theoretical coefficients for different models are shown in
Figure~\ref{fig:nukerstats1}. The circles and the squares denote
inefficient and efficient disc accretion respectively.  Diamonds
represent synchrotron X-rays from the jet.  Open and closed symbols
are for flat ($\alpha_{\rm R}=0$) and steep ($\alpha_{\rm R}=0.5$) 
spectrum sources, respectively.  
All model coefficients are calculated assuming $p=2$ (but see 
\S\ref{sec:jet_xrays} for a discussion of steeper 
electron distributions as due, for example, to the effects of cooling).
To highlight the sensitivity of the coefficients to variations in
$\alpha_{\rm R}$, we also plotted the tracks in $\xi_{\rm RX}-\xi_{\rm RM}$
space for changes in $\alpha_{\rm R}$ between 0 and 0.5.


For the models in which X-rays are produced by radiatively 
inefficient accretion,
we have $q=2.3$ and $\partial \ln\phi_{B}/\partial \ln \dot{m} = -
\partial \ln \phi_{B}/\partial \ln M= 0.5$ (see \S\ref{sec:adaf}).
The predicted coefficients for both flat and steep spectrum sources
lie roughly within the $3\sigma$ contours of the observed sample.
Thus, radiatively inefficient accretion is consistent with the observations. 
This conclusion is also true for the general class of models for which 
$q=2$.

For the standard disc--corona model, we have $q=1$ (high radiative
efficiency) and $\partial \ln \phi_{B}/\partial \ln \dot{m} = -
\partial \ln \phi_{B}/\partial \ln M= 0.5$ (see \S\ref{sec:corona}).
The model predictions fall well outside the $3\sigma$ contours of the
data and are therefore inconsistent with the observations.

For the models in which the X-rays are produced by optically thin
synchrotron emission from the jet itself, the flat spectrum model
predictions are marginally consistent with the data (the model point
is close to the $3\sigma$ contour), while the steep spectrum model
point lies well outside the $3\sigma$ contour.  

These results suggest that, in a statistical sense, the correlations
between radio luminosity, X-ray luminosity and mass of active black
holes require a radiatively inefficient accretion flow
coupled with a (scale invariant) synchrotron emitting jet. 
It is important to stress here that this result does not suggest that the
ADAF model for the accretion flow is the correct one, and convection
and/or powerful outflows are unimportant. In fact, the nature of the
radio emission and the mere existence of the fundamental plane are  
clear indications that the full dynamical model for the accretion flow should
at some level include, in a self-consistent manner, 
the effects of the backreaction from the
outflowing gas. However, our results rather 
dictates that the radiative efficiency of the flow {\it must} be low.
This conclusion holds over the observed range of $L_{\rm X}/L_{\rm Edd}$,
and therefore for sources which are substantially sub-Eddington. In 
\S\ref{sec:accretionmodechanges} we discuss what should be expected from
similar studies performed on samples that include a larger number of
very luminous black holes.  

The jet synchrotron model for the X-ray emission with $p=2$ is only
marginally consistent with the data. However, steeper electron
distributions ($p=3$, for example) do predict values for the
correlation coefficient that are well within our 3-$\sigma$ contours.
An assessment of the relevance of such models should be made taking in
due account the effect of cooling, as we discuss in \S\ref{sec:jet_xrays}.

\section{Discussion}
\label{sec:discussion}

The main result of our work is the discovery of a ``fundamental plane'' of
black hole activity. That is, if we define the instantaneous state of
activity of a black hole of mass $M$ (in units of solar masses), by the
radio luminosity at a fixed frequency (for example at 5 GHz), and by the
hard X-ray luminosity (for example in the 2-10 keV energy band) of its
compact core, and represent such an object as a point in the
three-dimensional space ($\log L_{\rm R},\log L_{\rm X},\log M$), all the
points representing black holes (either of stellar mass or supermassive)
will lie preferentially on a plane, described by equation (\ref{eq:fp}).

We note that in the case the radiatively inefficient disc model ($q=2.3$)
the fundamental plane equation (\ref{eq:fp}) implies that the radio
luminosity satisfies:
\begin{equation}
\label{eq:lr_bigmdot}
L_{{\rm R,}q=2.3} \propto \dot{m}^{1.38} M^{1.38} = \dot{M}^{1.38},
\end{equation}
i.e., $L_{\rm R}$ scales with the {\em physical accretion rate} only.  This
is very close to the predicted dependence of $L_{\rm R} \propto
\dot{M}^{1.42}$ from \ref{sec:sync} for the canonical parameter choice of
$\alpha=0$, $p=2$, and $\phi_{B}\propto M^{-1/2}\dot{m}^{1/2}$, for which
the kinetic jet power $W_{\rm jet}$ is directly proportional to the
physical accretion rate, $W_{\rm jet} \propto \dot{M}$ (see also Falcke \&
Biermann 1995). Then, the total power released by the accretion/jet 
system may be written as
$W_{\rm tot} \simeq \dot M c^2 = L_{\rm bol}+W_{\rm jet}+W_{\rm
  adv,conv}$, where the first term on the right hand side is the total
radiated luminosity and the last one include contributions from
the energy advected and/or stored in the convective motions.
Our results suggest that the flow must be radiatively inefficient,
therefore, for small enough accretion rates we have
 $L_{\rm bol}\simeq \dot M \dot m c^2 \ll \dot M c^2 \sim W_{\rm jet}+W_{\rm
  adv,conv}$. On the other hand,  $ W_{\rm jet} \propto W_{\rm
  adv,conv} \propto  \dot M c^2$. Therefore, the issue of what
the relative fraction of the total accretion energy dissipated into
the jet is (or, alternatively, of when a source is ``jet dominated'';
Fender, Gallo \& Jonker  2003; Falcke, K\"ording \& Markoff 2003) 
reduces to the
determination of the value of the constant $W_{\rm jet}/W_{\rm adv,conv}$.
This requires the knowledge of the structure functions $\psi$
in eq.~(\ref{eq:sc_inv}), i.e. the specification of a jet model or
the direct measure of the total kinetic power carried by the jet,
together with a dynamical model for the disc-jet coupling, which is
clearly beyond the scope of this paper.

\subsection{The thickness of the fundamental plane}
\label{sec:scatter}
The fundamental plane defined in eq.~(\ref{eq:fp}) is not, however, razor
thin. The sources are substantially scattered around it, with a dispersion
$\sigma=0.88$ in $\log L_{\rm R}$ (corresponding to a dispersion of
$\sigma_{\perp}=0.62$ perpendicular to the plane).  Such a scatter is not
at all surprising. Theoretically, it can be explained in large part by the
scatter in the radio spectral index $\alpha_{\rm R}$.  For example, if we
consider the ADAF ($q=2.3$) model with fixed $p=2$, we expect the
theoretical relationship $L_{\rm R}=(0.62+0.29\alpha_{\rm R})\log L_{\rm
X}+(0.80-0.62\alpha_{\rm R})\log M +b_{\rm R}$. For the sake of simplicity,
let us assume now that the unknown radio spectral index of all the observed
sources is normally distributed, with $\alpha_{\rm R}=\langle \alpha_{\rm
R} \rangle \pm \sigma_{\alpha}$. In order to fit the observed data, we need
$\langle \alpha_{\rm R} \rangle \approx 0$ and $b_{\rm R} \approx 7$.  The
dispersion in the radio luminosity due to the dispersion in the unknown
parameter $\alpha_{\rm R}$ is then $\sigma_{\rm R} \ga
\sigma_{\alpha}\sqrt{0.33^2 \left(\Delta
\log{M}\right)^2+0.29^2\left(\Delta \log{L_{\rm X}/M}\right)^2} \sim 2
\sigma_{\alpha}$ where $2 \Delta \log{M}$ is the range in $M$ and $2 \Delta
\log{L_{\rm X}/M}$ the range in luminosity spanned by our sample. Thus, a
dispersion in the unknown radio spectral index of $\approx 0.3$ (consistent
with the observations), can give a large contribution to the observed
intrinsic scatter.

In addition to the scatter produced by the diversity in $\alpha_{\rm
  R}$ (which introduces scatter in the fundamental plane relation
simply by the fact the {\em orientation} of the plane is sliiightly
different for different $\alpha_{\rm R}$) the {\em intercept} $b_{\rm
  R}$ of the plane also varies as we look at different jet models and
different spectral indices.  It is not clear a-priori how much it does
so, as this depends on how the shape functions $\psi_f$ (see
section~\ref{sec:sync}) vary in order to produce the different
$\alpha_{\rm R}$.  While for one spectral class with uniform
$\alpha_{\rm R}$ the normalization cancels out when determining the
correlation coefficients, this is not true when comparing jets from
different spectral classes, i.e., with varying $\alpha_{\rm R}$.  The
fact that the scatter in the relation is moderate indicates, however,
that the change in $\psi_f$ over the range of radio spectral slopes
that enter our sample is also moderate.  

Ideally, we should therefore restrict the above analysis to
sub-classes of sources which follow a very narrow range in
$\alpha_{\rm R}$.  However, at the current stage the data do not allow
such a division, both because the sample is too small and because the
spectral indices are often not known to an accuracy that would allow
such a treatment.  The rough division of our sample into flat and
steep spectrum sources nonetheless shows that such a treatment is
possible and does lead to a reduction in scatter in the well defined
and well sampled class of flat spectrum sources, for which we can be
confident that the radio emission does indeed originate in the
self-absorbed core of the jet (a prerequisite of the scale invariance
model by Heinz \& Sunyaev, 2003).

A major source of scatter could be relativistic beaming, that we
discuss in the following section. 
Also to be taken into account is the influence of the black hole
spin on jet formation \cite{bz77,koi02}.  
While we still expect the accretion disc to
have significant influence on the jet power, as the magnetic field
necessary to tap the black hole rotational energy must be provided
and/or confined by the accretion disc (thus setting the field strength
and the jet power; see Livio, Ogilvie \& Pringle 1999; Meier 2001), 
the spin itself will enter as a sensitive
parameter into the radio luminosity as well, introducing scatter in
any correlation.  Once again, strong future constraints on the
tightness of this relation may help constrain the possible influence
of black hole spin (see e.g. Fender 2003, for a discussion of the GBH case).

\subsection{Relativistic beaming and selection effects}
\label{sec:beaming}

Because the jets that dominate radio emission from active black holes
are known to be relativistic, Doppler boosting of synchrotron
radiation (relativistic beaming) must be taken into account when
studying these sources. 

For the purpose of our study it is of primary importance to avoid
strongly beamed sources, whose intrinsic (unbeamed) radio luminosity
can be difficult to infer. This is the reason why BL Lac object have
been excluded from our study. Furthermore, it is also important to
take into account possible strong selection effects 
due to relativistic beaming, that would skew the observed correlations. 
Then, in a sample of randomly oriented jetted sources (but without
the aligned ones), relativistic beaming should be an additional source
of scatter in any correlation involving radio luminosities, the
magnitude of which depends on the average relativistic speed of the jets.

For the SMBH in our sample,  the first selection criterion is the
availability of a mass measurement of the central black hole. 
Thus, the main selection biases do not come from
flux limits, and no preferential selection of bright (beamed) sources
should be operating, although anti-selection effect might be present. 

Possible contaminations from
beamed sources, however, could still be present among flat spectrum
nearby LLAGN or type 1 Seyfert (we have 27 of them in the sample), 
given the current uncertainties on the exact
nature of their radio emission. 
Nevertheless, there are many reasons
to believe that these sources are not biased towards jet axes close to the
line of sight. Flat spectrum LLAGN were originally selected in the optical
band, from the Palomar spectroscopic survey \cite{hfs97a}. Many of those
type 1 low-luminosity Seyfert and LINERS do indeed possess double-peaked
broad emission lines \cite{hor00}, clearly indicating a non negligible
inclination angle.  The radio properties of a number of these 
sources were also studied by Nagar et al. (2002). 
 They found that LLAGN lie on the extrapolation of
the FR I/FR II correlations in the $L_{\rm R} - L_{H\alpha}$ plane, while a
substantial offset would be expected if they were all strongly affected by
beaming.
Once again, we are led to the conclusion that nearby LLAGN and Seyfert
1 included in our sample have a random orientation angle, that  
should disfavor strongly beamed
sources. Finally we note that even if a few of these source were
indeed strongly Doppler boosted, their small number will not affect
significantly the measured value of a correlation coefficient
calculated in logarithmic space (logarithmic weighting).

Selection effects due to beaming in Galactic sources may be more
subtle. Due to the overall scaling of radio luminosity with mass, 
GBH are expected to be on average more  radio quiet than SMBH (see
also \S\ref{sec:accretionmodechanges}). It is indeed well known that
galactic X-ray binaries are  difficult to detect in the radio band,
and this might already determine a selection bias towards beamed
sources.  
In particular, sources selected from GBI monitoring (see \S\ref{sec:sam_gbh})
might be more beamed. If indeed GBH were all affected by strong
selection biases toward beamed emission, the derived correlation
coefficients of the fundamental plane, $\xi_{\rm RM}$ in particular,
might be incorrect. However, we know from observation of proper motion
during radio outbursts  that the jets of the classical
microquasars GRO J1655-40 and GRS 1915+105 have large angle with the
line of sight (see e.g. Fender 2003). Furthermore, 
for all the other GBH in our sample, apart from
LS5039, orbital parameters have been inferred  from optical
spectroscopy, and in none of them the inferred inclination  angle
appears to be small.  Therefore, unless the black hole jets are all
misaligned with the orbital angular momentum \cite{mac02}, we can
exclude that strong beaming affects our sample.

A similar conclusion had already been reached by an accurate study of
the observed scatter ($\sim 0.7$) 
about the $L_{\rm R} - L_{\rm X}$ correlation in a larger
sample of low/hard state GBH \cite{gfp03}. An analogous argument could
be put forward for our entire sample, with the similar conclusion
(given the similar amount of scatter found here) that the maximum 
Doppler factor should not be too large. This conclusion would be
further strengthen by noting that the observed scatter 
about the fundamental plane is not much larger
than what expected from the intrinsic scatter in the radio spectral index
$\alpha_{\rm R}$ (see \S\ref{sec:scatter}). 
However, precise constraints on the Doppler factor could only 
be obtained by analyzing more carefully selected samples with
well determined (possibly fixed) radio spectral indices.

\subsection{Steep spectrum sources}
\label{sec:steepspectra}
For the steep spectrum sources in our sample, the origin of the radio
emission is not entirely clear.  Steep spectra in general imply optically
thin emission.  This could be an indication that the emission is dominated
by regions far away from the central engine, e.g. in the diffuse extended
radio lobes or hot spots.  This would prohibit any treatment that makes use
of the scale invariant model, though scaling arguments for the large scale
emission still apply \cite{heinz02}.  Steep spectrum sources should then be
excluded from the sample completely, and the fact that they still fit into
the plane would then mostly be due to selection effects.  It should be
noted that in this case the radio luminosity should strongly correlate with
the black hole mass (which determines the kinetic jet power $W_{\rm jet}$
and thus the large scale radio luminosity), while it should not be related
to the instantaneous X-ray flux, because the radio lobes reflect the mean
power output by the central source, averaging out its temporal variability.
We expect this to be a problem for Galactic black hole sources, where we
have lower resolving power.  For supermassive black holes and for nearby
LLAGN in particular, we can be more confident that the large scale emission
is well resolved and thus does not contribute to the core emission used in
our sample.

It is also possible that some of the steep spectrum sources can be
associated to core jet emission which is optically thin at radio
frequencies, i.e., their self-absorption break lies below 5 GHz.
Because we would expect the self-absorption break frequency to be
lower for higher $M$ and lower accretion rates $\dot{m}$ \cite{hs03},
this effect would imply that steep spectra should predominantly be
observed in SMBH which are relatively X-ray dim.  While the trend with
$L_{\rm X}$ is not clear in the data, the fact that the steep spectrum
sources in our sample almost exclusively belong to the SMBH class does
suggest that at least some of the sources in the sample might indeed
be core dominated steep spectrum sources, for which the theoretical
analysis of \S\ref{sec:physics} holds.

\subsection{X-ray emission from jets}
\label{sec:jet_xrays}
Our analysis, taken at face value, does suggest that accretion models
for the origin of the X-ray emission fit the data somewhat better than
pure jet models.  However, at X-ray energies the effects of cooling on
the particle spectrum cannot be ignored.  In the context of scale
invariant models, taking cooling into account is not possible.  Proper
treatment of particle transport including the effects of cooling may
lead to different results for the synchrotron X-ray model.  In fact,
simply using $p=3$ in eq.~(\ref{eq:tildexi}) 
(as expected in a continuous particle injection
model with cooling) moves the flat spectrum point close to the
$1\sigma$ contour of the observed correlation in Fig.~\ref{fig:nukerstats1}.

For the case of the galactic black hole GX 339-4 in the hard state, where
the radio--X-rays correlation is well established, Markoff et al. (2003)
were able to fit the observed slope ($\xi _{\rm RX,\, 339} \simeq 0.71$)
with a synchrotron X-ray model assuming $p=2.15$ (and thus $\alpha_{\rm
X}=0.58$, which is also close to the average slope of the X-ray spectral
index of the source in the hard state). Here we have found that, in a
statistical sense, higher values of $p$ are  needed to explain the
observed correlation for the entire sample in the framework of the jet
synchrotron model for the X-ray emission.

Furthermore, for $p=2$ the X-ray synchrotron luminosity from the jet scales
like $L_{\rm X,synch} \propto M^{1.25}$, while the disc X-ray luminosity
follows $L_{\rm X,disc} \propto M$.  Thus, {\em if} the X-ray emission in
GBH jets like GX 339-4 were indeed due to synchrotron emission, we would
expect an even stronger contribution from synchrotron X-rays relative to
the disc emission in AGNs with the same accretion rate, where the general
wisdom (mainly based on accurate analysis of detailed X-ray spectra) is
still that the X-rays originate predominantly in the disc.  For $p=3$, on
the other hand, the synchrotron X-rays follow $L_{\rm X,synch} \propto M$,
and the relative contribution from jet and disc will be independent of $M$.

Finally, the X-ray spectral variability behaviour of both GBH and Seyfert
galaxies is consistent with pivoting of the spectrum at X-ray energies. If
the X-ray emission is produced by synchrotron radiation and there is no
cooling break at longer wavelengths, such a power-law variability would
imply a radio variability which is orders of magnitude larger than observed
(see discussion in Zdziarski et al. 2003).

It seems, therefore, that {\em if} the X-ray emission in some of the
sources is dominated by synchrotron emission from the jet, the effects of
cooling must play an important role in determining the observed
correlation.  In the context of the scale-invariant model, it is rather
problematic to take cooling into account; this is why a theoretical
assessment of the role of the cooling on the observed correlation is beyond
the scope of this paper.

\subsection{Estimating black holes masses and the nature of ULXs}
Due to the intrinsic scatter, the predictive power of the $\log M - \log
L_{\rm X} - \log L_{\rm R}$ relation is currently limited in scope.
Nonetheless, from the fundamental plane equation (\ref{eq:fp}), we can
derive a formal relationship between the observed fluxes in the radio (at
5 GHz) and in the X-ray (2-10 keV) band (in erg s$^{-1}$ cm$^{-2}$), the
distance to a source $D$ (in Mpc) and its unknown mass. We obtain
\begin{equation}
\log M \simeq 16.3 + \log D + 1.28 (\log F_{\rm R} - 0.60 \log F_{\rm
  X}) \pm 1.06.
\end{equation}

In particular, we can predict in which region of $\log L_{\rm R}-\log
L_{\rm X}$ space we would expect to find intermediate mass black holes, if
they exist, and specifically, where we would expect Ultra-Luminous X-ray
sources (ULX; Makishima et al. 2000) based on their X-ray flux, should they
be intermediate mass black holes, as opposed to intrinsically beamed low
mass black holes.

A clear prediction of the fundamental plane equation in this sense is
that, if ULX are stellar mass objects (with $M<30$), 
and their radio emission is not strongly Doppler boosted, there is an
upper limit to their radio flux at 5 GHz. For example, a
ULX with and X-ray luminosity (in the 2-10 keV band) of 10$^{40}$ ergs
s$^{-1}$ at a distance of 10 Mpc will have
\begin{equation}
\log (F_{\rm R}/{\rm Jy}) \la -6.5 -2 \log (D/10) +0.6 (\log L_{\rm
  X}/40) \pm 0.9
\end{equation} 
so its flux will be at most
few microJansky. 

As an example, we refer to the recent claim of the observation of
synchrotron radio emission from the ULX 2E 1400.2-4108, in the dwarf
irregular galaxy NGC 5408 (at a distance of $\sim 4.8$ Mpc)
\cite{kaa03}. The observed radio (0.26 mJy at 4.8 GHz) and X-ray fluxes
($2.7 \times 10^{-12}$ erg s$^{-1}$ cm$^{-2}$ in the 0.3-8 keV band), if
indeed both associated with a compact source at the distance of NGC 5408,
would imply a mass of $\log M \simeq 4 \pm 1$.  However, we should stress
here that such an estimate only holds {\em if the radio emission is not
beamed}.  The uncomfortably high value for the estimated mass, the steep
inferred radio spectral index and the radio to X-ray flux ratio, all
suggest that the observed intensity of the radio emission in this
source may be Doppler
boosted by a relativistic jet pointing in our direction as already
suggested by Kaaret et al. (2003).
 
\subsection{Accretion mode changes}
\label{sec:accretionmodechanges}
It is well accepted, both from theory and observations, that accretion
can proceed in different modes (or states), with different radiative
efficiencies and spectral properties \cite{abr98,don02,fkr02,nar02},
mainly driven by variations of the dimensionless accretion rate $\dot
m$.  Here we have demonstrated (\S\ref{sec:physics}) that the
correlation between radio and X-ray luminosity in GBHs and SMBHs can
provide valuable constraints on the emission mechanisms in these bands
and on the physics of accretion. We have shown how low radiative
efficiency accretion models can possibly provide the underlying
scalings that most easily explain the properties of the observed
fundamental plane.  Radiatively efficient thin disc--corona systems
are clearly inconsistent with the data. The only possibility for these
models to be reconciled with our observations would entail a highly
radiatively inefficient corona, where most of the dissipated magnetic
energy is converted into bulk kinetic energy of outflowing gas
\cite{bel99}, as described by Merloni \& Fabian (2002). 
This would require some specific model for the magnetic 
dissipation processes, and for their
scaling with $M$ and $\dot m$, which are currently unknown.

However, because both such modes of accretion are expected to occur
only below accretion rates $\la$ few percent of Eddington, we would
expect the $\log M - \log L_{\rm R} - \log L_{\rm X}$ correlation to
break down at around this limit. In other words, we should expect the QSOs
and the bright Seyferts in our sample, which occupy the region of high
accretion rates and are independently known to have spectral
characteristics inconsistent with models of low radiative efficiency,
to depart from the observed correlations. Also in GBHs, 
it has indeed been shown that the correlation between radio
and X-ray luminosity breaks down as the sources switch to their high
states \cite{mac03,gfp03}. In particular, high luminosity states of
GBH (high or very high states) show highly variable radio fluxes and
radio to X-ray ratios \cite{mr94,fen99}. If this reflects a general
property of the disc--jet coupling at high accretion rates, a
substantial increase in the scatter should appear above the critical
accretion rate.

\begin{figure}
\psfig{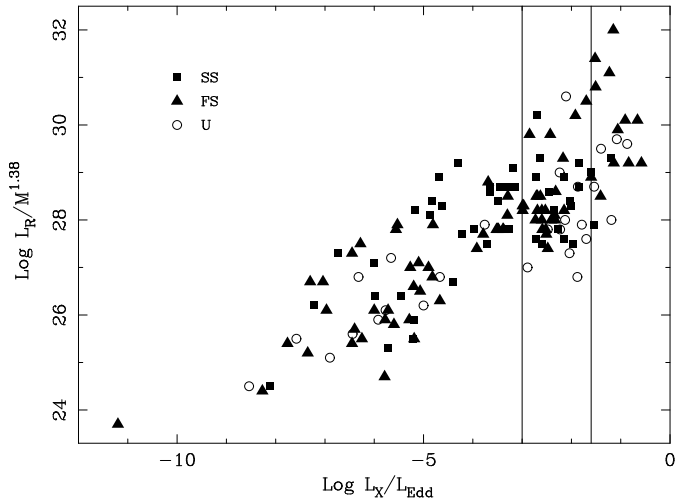}
\caption{The radio luminosity $\log L_{\rm R}$, divided by $M^{1.38}$ as a
function of the ratio $L_{\rm X}/L_{\rm Edd}$. Solid squares are for steep
spectrum sources, solid triangles for flat spectrum ones and open circles
for sources with undetermined radio spectral index. Two vertical lines mark
the boundary of the region where we expect the critical luminosity for the
mode change between radiatively inefficient and efficient accretion.  The
scaling for the radio luminosity with mass is obtained directly from the
fundamental plane equation (\ref{eq:fp}) by imposing that the X-ray
luminosity scales linearly with black hole mass.}
\label{fig:mdotcr}
\end{figure}

In order to illustrate this
point further, in Figure~\ref{fig:mdotcr} we plot, as a function of
the ratio $L_{\rm X}/L_{\rm Edd}$, the radio luminosity divided by
$M^{1.38}$ (such a scaling for the radio luminosity with mass is
obtained directly from the fundamental plane equation (\ref{eq:fp}) by
imposing that the X-ray luminosity scales linearly with black hole
mass).  As indeed expected, by rescaling the radio luminosity in such
a way all the different tracks corresponding to different mass bins in
Fig.~\ref{fig:lr_lx_bin}b collapse into a single one (with some
residual scatter).  The region between the two vertical lines
corresponds to the expected values of $L_{\rm X}/L_{\rm Edd}$ above
which a change of accretion mode, from radiatively inefficient to
standard radiatively efficient is expected to occur.
Our SMBH sample is still limited in order to test
whether such a change is observed. In fact, we select out of our
sample the majority of bright quasars, both radio quiet and radio
loud, because of the lack of a reliable mass estimates for these
sources.  However, there is indeed a hint of 
an apparent increase in the scatter about the
correlation, above the critical accretion rates, 
as seen in Figure~\ref{fig:mdotcr}.

We may thus speculate that the famous (and still much debated, see
e.g. Cirasuolo et al. 2003) radio loud/radio quiet dichotomy of
quasars will appear only at the highest values of $\dot m$, and be
caused mainly by a switch of accretion mode analogous to the high/very
high transition in GBH.  At low accretion rates, black holes seem to
follow the more regular behavior circumscribed by the fundamental
plane of eq.~(\ref{eq:fp}). Such sources not only tend to be radio
loud \cite{hp01,ho02}, but also their radio loudness parameter,
$R_{\rm X}$ (here defined as the ratio of radio to X-ray luminosity),
obeys the following scaling: $R_{\rm X}\equiv L_{\rm R}/L_{\rm X}
\propto L_{\rm X}^{-2/5} M^{4/5}$. Therefore, the smallest the 
X-ray luminosity, the more radio loud these sources are \cite{ho02}. 
In this regime, no dichotomy has to be expected, as
already suggested by Nagar et al. (2002).

\section{Summary}
\label{sec:summary}

We have shown how the coupling between accretion flows and jet emission in
black holes can be understood by studying large samples of {\it both}
galactic (GBH) and supermassive (SMBH) black holes with measured mass,
observed at both radio and X-ray wavelengths.

We have compiled, from the existing literature, a sample of $\sim$100
supermassive black holes and $\sim$50 different observations of galactic,
stellar mass black holes. We required that the SMBH have a reliable
measure of the central black hole mass, either (preferentially) direct, via
high resolution kinematics studies of surrounding stars and gas or
reverberation mapping of the broad line region (for luminous Seyfert 1
galaxies and Quasars), or, indirectly, thanks to the empirical correlation
between black hole masses and central velocity dispersion. Although such
selection criteria introduce a number of biases in the final distribution
of $M$, which are difficult to control, our final sample
spans a large enough range in both mass and accretion rate to unveil major
trends in the physical properties of active black holes. This is a
crucial point, and represents the main novelty of our approach. Most
previous studies on the subject have only considered correlations of jet 
(radio) luminosity with black hole mass or with luminosity in other wavewbands
(and hence with accretion rate) separately.

We have performed a partial correlation analysis on the sample, and
concluded that the radio luminosity is strongly correlated with {\em both}
black hole mass {\em and} X-ray luminosity, while in turn, the X-ray
luminosity correlates with both mass and radio luminosity. In light of
this, we have carried out a multivariate 
linear regression analysis and derived the best
fit correlation coefficient between radio luminosity, X-ray luminosity and
black hole mass.

The data from the entire sample allow us to put tight constraints on the
correlation coefficients. Remarkably, we find that the sources lie
preferentially on a plane (``the fundamental plane of black hole
activity'')
in the three dimensional $(\log L_{\rm R},\log
L_{\rm X},\log M)$ space, described by the equation $\log L_{\rm
R}=(0.60^{+0.11}_{-0.11}) \log L_{\rm X} +(0.78^{+0.11}_{-0.09}) \log M +
7.33^{+4.05}_{-4.07}$. The scatter around this plane is, however,
significant ($\sigma_{\rm R}=0.88$).

From a theoretical point of view, these results clearly suggest that the
{\em ansatz} of scale invariance for the disc--jet coupling \cite{hs03}
captures the main physical properties of such systems. Thus, a universal
theoretical 
scaling between the radio flux at a given frequency and both mass and
accretion rate can be derived, {\it independently of the jet model}, with
scaling indices that depend only on the (observable) spectral slope of the
synchrotron emission in the radio band, and on the accretion mode. Also, it
is possible to predict the correct amount of scatter for any such
relationship.

By comparing the observationally derived correlation coefficients to the
theoretically predicted ones, we are able to put constraints on accretion
models and on the disc--jet coupling. We demonstrate that the X-ray emission
from black holes accreting at less than a few per cent
 of the Eddington rate cannot
be produced by radiatively efficient accretion, while
radiatively inefficient accretion flows agree well with the data.
The optically thin X-ray jet synchrotron emission model is only
marginally consistent with the observed correlation; 
however, cooling of the electrons should be properly taken into account
before drawing a firmer conclusion on the issue.

The fundamental plane equation also implies that low luminosity black
holes should be more and more radio loud as the accretion rate
decreases, as observed \cite{ho02}. The possible dichotomy between
radio quiet and radio loud Quasars could instead be due to a switch of
accretion mode at the highest accretion rates. 
Therefore, 
more accurately selected samples of more luminous sources, including a
significant number of both radio quiet and loud quasars with well
determined radio spectral properties, should allow the observation of 
bends, discontinuities or even bifurcations of the fundamental plane 
associated with accretion mode switches and with the turning on and
off of the most powerful radio sources.

\section*{Note added in proofs}
After the submission of this manuscript another work has appeared as
preprint \cite{fkm03}, that is similar in scope, and reaches very similar
conclusions about the scale invariant properties of low luminosity
black holes.

\section*{Acknowledgments}
We are grateful to Marat Gilfanov for his useful comments.
We thank Eric Feigelson and the Statistical Consulting Center for Astronomy
(SDASS) for providing the code for the partial correlation analysis and
some advice on its usage. This research has made use of the NASA/IPAC
extragalactic database (NED), which is operated by the Jet Propulsion
Laboratory, Caltech, under contract with NASA. We have also made use of
LEDA, supplied by the LEDA team at the CRAL-Observaoire de Lyon.  This
research has made use of the publicly available radio monitoring data
from the Green Bank Interferometer.  The Green Bank Interferometer is a
facility of the National Science Foundation operated by the NRAO in support
of NASA High Energy Astrophysics programs.  This research made use of
results provided by the ASM/RXTE teams at MIT and at the RXTE SOF and GOF
at NASA's GSFC.

\bsp

\label{lastpage}

\end{document}